\documentclass[11pt]{article}

\pdfoutput=1

\usepackage{epsf}
\usepackage{cite}
\usepackage{color}
\usepackage{graphicx}
\usepackage{amssymb,amsmath}

\usepackage{hyperref}
\usepackage{amsmath}
\usepackage{amssymb}
\usepackage{epsfig}
\usepackage{cite}
\usepackage{color,colordvi}

\newcommand{\gsim}{ \mathop{}_{\textstyle \sim}^{\textstyle >} }

\newcommand{\beq}{\begin{equation}}
\newcommand{\eeq}{\end{equation}}

\newcommand{\One}{\hbox{1\kern-.24em I}}

\newcommand{\lascia}[1]{}
\makeatletter
\def\art{\@ifnextchar[{\eart}{\oart}}
\def\eart[#1]#2#3#4#5#6{{\rm #2}, {#3 #4} {\rm (#6) #5} [arXiv:{\hhref{#1}}]}
\def\hepart[#1]#2{{\rm #2, arXiv:\hhref{#1}}}
\newcommand{\oart}[5]{{\rm #1}, {#2 #3} {\rm (#5) #4}}

\def\circa#1{\,\raise.3ex\hbox{$#1$\kern-.75em\lower1ex\hbox{$\sim$}}\,}
\def\mathscr#1{{\fam\rsfsfam\relax#1}}

\newcommand{ \be}{\begin{equation}}
\newcommand{ \ee}{\end{equation}}
\newcommand{ \bea}{\begin{eqnarray}}
\newcommand{ \eea}{\end{eqnarray}}
\newcommand{ \mysmall}[1]{\scriptscriptstyle #1} 

\newcommand{ \gmt}   {$g$$-$$2$~}
\def\eq#1{eq.~(\ref{#1})}

\begin{document}
\begin{titlepage}

\begin{center}
\hfill CERN-PH-TH/2012-017
\vskip 0.4 cm
\vspace{1cm}
{\Large\bf Testing new physics with the electron \boldmath $g-2$ \unboldmath}
\vspace{0.2cm}
\end{center}

\begin{center}
{\bf G.\ F.\ Giudice}$^{a}$,
{\bf P.\ Paradisi}$^{a}$, 
{\bf M.\  Passera}$^{b}$

	\vspace{0.4cm}

{\em $^a$CERN, Theory Division, 1211 Geneva 23, Switzerland}\\
{\em $^b$Istituto Nazionale Fisica Nucleare, 35131 Padova, Italy}
\vspace{1cm}

\today

\end{center}

\vskip .2in
\setlength{\baselineskip}{0.2in}

\centerline{\bf Abstract}
\vskip .1in
\noindent

We argue that the anomalous magnetic moment of the electron ($a_e$) can be used to probe new physics. We show that the present bound on new-physics contributions to $a_e$ is $8\times 10^{-13}$, but the sensitivity can be improved by about an order of magnitude with new measurements of $a_e$ and more refined determinations of $\alpha$ in atomic-physics experiments. Tests on new-physics effects in $a_e$ can play a crucial role in the interpretation of the observed discrepancy in the anomalous magnetic moment of the muon ($a_\mu$). In a large class of models, new contributions to magnetic moments scale with the square of lepton masses and thus the anomaly in $a_\mu$ suggests a new-physics effect in $a_e$ of $(0.7\pm 0.2)\times 10^{-13}$. We also present examples of new-physics theories in which this scaling is violated and larger effects in $a_e$ are expected. In such models the value of $a_e$ is correlated with specific predictions for processes with violation of lepton number or lepton universality, and with the electric dipole moment of the electron.  

\end{titlepage}

\tableofcontents
\newpage

\section{Introduction}
\label{sec:1}

The anomalous magnetic moment of the muon $a_\mu = (g-2)_\mu /2$ is one of the most celebrated tests of the Standard Model {\small (SM)}. Indeed, the high precision of its theoretical and experimental determinations makes $a_\mu$ a powerful test on new physics~\cite{gm2_rev_NP1}. The situation has become especially intriguing with the $\sim 3.5 \sigma$ reported discrepancy between the {\small SM} prediction and the experimental value~\cite{bnl, Jegerlehner:2009ry, HLMNT11, DHMZ11}
\be
\Delta a_\mu =a_\mu^{\mysmall \rm EXP}-a_\mu^{\mysmall \rm SM} = 2.90 \, (90) \times 10^{-9}.
\label{eq:gmu}
\ee

On the other hand, the anomalous magnetic moment of the electron $a_e$ has never played a role in testing ideas beyond the {\small SM}. In fact, it is believed that new-physics contaminations of $a_e$ are too small to be relevant and, with this assumption, the measurement of $a_e$ is employed to determine the value of the fine-structure constant $\alpha$.

The aim of this paper is to emphasize that the situation has now changed, thanks to advancements both on the theoretical and experimental sides. Indeed, the theoretical prediction of $a_e$ has been refined to an unprecedented accuracy and its experimental value is now known with smaller errors. At the same time, good determinations of the fine-structure constant have been obtained from atomic physics experiments, providing a value of $\alpha$ that is completely independent of the measurements of $a_e$. As a result, $a_e$ can now be viewed as a very useful probe of physics beyond the {\small SM} and the situation is going to become even more promising soon, as efforts are underway to reduce significantly both theoretical and experimental errors. The most exciting aspect of the story is that $a_e$ will soon provide us with a crucial consistency check of new-physics interpretations of the alleged discrepancy in $a_\mu$. Moreover, in certain classes of models, if $\Delta a_\mu$ is caused by new physics, then it is possible to correlate the value of $a_e$ with various other rare processes violating lepton universality or individual lepton number.

The paper is organized as follows. In sect.~\ref{sec:2} we carefully review the present status of the {\small SM} prediction of $a_e$ and confront it with the experimental measurement. In sect.~\ref{sec:3} we discuss future improvements in the theoretical and experimental results
and show their impact for probing new physics. We use the prototype examples  of supersymmetry (sect.~\ref{sec:susy}), of a light pseudoscalar (sect.~\ref{sec:ps}), and of vector-like fermions (sect.~\ref{sec:vec}) to illustrate generic features of theories beyond the {\small SM}. Our results are summarized in sect.~\ref{sec:conc}.

\section{Status of the electron $g-2$}
\label{sec:2}

\subsection{The experimental situation}

The classic series of measurements of the electron and positron anomalous magnetic moments carried out at the University of Washington yielded in 1987 the value $a_{e}^{\mysmall \rm EXP} = 115 \, 965 \, 218 \, 83 \, (42) \times 10^{-13}$~\cite{VanDyck:1987ay,Mohr:2000ie}. More recently, a new determination of the electron \gmt has been performed by Gabrielse and his collaborators at Harvard University, with the result~\cite{Gabrielse}
\be 
a_{e}^{\mysmall \rm EXP} = 115 \, 965 \, 218\, 07.3 \, (2.8) \times 10^{-13}.
\label{eq:aEEX}
\ee
The uncertainty of this result, $\delta a_{e}^{\mysmall \rm EXP} =2.8 \times 10^{-13}$, {\it i.e.}\ 0.24 parts in a billion (ppb), is 15 times smaller than that reported back in 1987. The two measurements differ by 1.8 standard deviations.

\subsection{The Standard Model prediction}

The {\small SM} prediction $a_{e}^{\mysmall \rm SM}$ is usually split into three parts: {\small QED}, electroweak ({\small EW}) and hadronic. Here we provide a summary 
of the present status of these contributions.

\subsubsection{QED contribution}

The {\small QED} term $a_e^{\mysmall \rm QED}$ arises from the subset of {\small SM} diagrams containing only leptons and photons. This dimensionless quantity can be cast in the general form~\cite{KM90}
\be
    a_e^{\mysmall \rm QED} \!\!=\! A_1 \!+ 
                   A_2 \biggl( \frac{m_e}{m_{\mu}} \biggr)  + 
                   A_2 \biggl( \frac{m_e}{m_{\tau}} \biggr)  + 
                   A_3 \biggl( \frac{m_e}{m_{\mu}},
                   \frac{m_e}{m_{\tau}} \biggr),    
\label{eq:amuqedgeneral}
\ee
where $m_e$, $m_{\mu}$ and $m_{\tau}$ are the electron, muon and tau lepton masses. The term $A_1$, arising from diagrams containing only photons and electrons, is mass and flavor independent.  In contrast, the terms $A_2$ and $A_3$, generated by graphs containing also muons and taus, are functions of the indicated mass ratios. The muon contribution to $a_e^{\mysmall \rm QED}$, although suppressed by $m_e^2/m_{\mu}^2  \sim 2.34 \times 10^{-5}$, is about ten times larger than the experimental uncertainty in \eq{eq:aEEX} (the tau contribution, suppressed by $m_e^2/m_{\tau}^2$, is of order $10^{-14}$). The functions $A_i$ ($i\!=\!1,2,3$) can be expanded as power series in $\alpha/\pi$ and computed order-by-order:
$
    A_i  = A_i^{(2)} \left(\alpha/\pi \right)
           + A_i^{(4)} \left(\alpha/\pi \right)^{2}
           + A_i^{(6)} \left(\alpha/\pi \right)^{3} +\cdots.
$
%


Only one diagram is involved in the evaluation of the one-loop (first-order in $\alpha$, second-order in the electric charge) contribution; it provides the famous result by Schwinger~\cite{Schwinger:1948iu}
\be
C_1 = A_1^{(2)} =  1/2.
\ee
Seven two-loop diagrams contribute to the fourth-order coefficient $A_1^{(4)},$
one to $A_2^{(4)}(m_e/m_{\mu})$ and one to $A_2^{(4)}(m_e/m_{\tau})$, while $A_3^{(4)}(m_e/m_{\mu},m_e/m_{\tau}) = 0$. The exact mass-independent coefficient
has been known for more than fifty years,
$
    A_1^{(4)} =  197/144 + \pi^2/12 + 3\zeta(3)/4 - (\pi^2/2) \ln2
    =    -0.328 \, 478 \, 965 \, 579 \, 193 \, 78 ...
$~\cite{So57-58-Pe57-58}, 
where $\zeta(s)$ is the Riemann zeta function.
The numerical evaluation of the exact expression for the mass-dependent coefficient $A_2^{(4)}$~\cite{El66,Passera:2004bj} with the latest {\small CODATA} recommended
mass ratios 
$ m_e/m_{\mu} = 4.836 \, 331 \, 66 (12) \times 10^{-3}$
and 
$ m_e/m_{\tau} = 2.875 \, 92 (26) \times 10^{-4}$
yields
$
     A_2^{(4)} (m_e/m_{\mu})  =  5.197 \, 386 \, 68 \, (26) \times 10^{-7} 
$
and
$
     A_2^{(4)} (m_e/m_{\tau})  =  1.837 \, 98 \, (33) \times 10^{-9}
$~\cite{Mohr:2012tt}.
The tiny errors are due to the uncertainties of the mass ratios. The sum of the above values provides the two-loop {\small QED} coefficient
\be
    C_2  =  A_1^{(4)} + A_2^{(4)}(m_e/m_{\mu}) + 
                  A_2^{(4)}(m_e/m_{\tau}) 
              =  - 0.328 \, 478 \, 444 \, 002 \, 55 \, (33).
\ee
The standard error $\delta C_2 = 3.3 \times 10^{-13}$ leads to a totally negligible $O(10^{-18})$ uncertainty in $a_e^{\mysmall \rm QED}$.


More than one hundred diagrams are involved in the evaluation of the three-loop (sixth-order) {\small QED} contribution. The exact result for the coefficient $A_1^{(6)}\!$, mainly due to Remiddi and his collaborators~\cite{Laporta:1996mq}, yields the numerical value
$
     A_1^{(6)} = 1.181 \, 241 \, 456 \, 587...
$ .
The analytic calculation of the mass-dependent coefficient $A_2^{(6)}(r)$ for arbitrary values of the mass ratio $r$ was completed in 1993 by Laporta and Remiddi~\cite{LR93}.  The exact formula contains hundreds of polylogarithmic functions, including harmonic polylogarithms~\cite{HPL}. Numerically evaluating these analytic expressions with the latest {\small CODATA} mass ratios given above, we obtain
$
     A_2^{(6)}(m_e/m_{\mu})  =  -7.373 \, 941 \, 62 \,(27)  \times 10^{-6}
$
and
$
     A_2^{(6)}(m_e/m_{\tau}) = -6.5830 \,(11)   \times 10^{-8}.
$
The same values can be obtained with the simple series expansions of~\cite{Passera:2006gc}, thus avoiding the complexities of these numerical evaluations. The contribution to $a_{e}^{\mysmall \rm QED}$ of these three-loop mass-dependent coefficients is  $-0.9 \times 10^{-13}$, i.e.\ about a third of the present experimental uncertainty $\delta a_{e}^{\mysmall \rm EXP}$, see \eq{eq:aEEX}.
The contribution of the three-loop diagrams with both muon and tau loop insertions in the photon propagator can be calculated numerically from the integral expressions of~\cite{Samuel91}. We get
$
     A_3^{(6)}(m_e/m_{\mu},m_e/m_{\tau}) = 1.909 \, 82 \,(34) \times 10^{-13},
$
a totally negligible $O(10^{-21})$ contribution to $a_{e}^{\mysmall \rm
QED}$.  
Adding up all the above values we obtain the three-loop {\small QED} coefficient
\be
    C_3    =   1.181 \, 234 \, 016 \, 816 \,(11).
\label{eq:C3}
\ee
The error $\delta C_3 = 1.1  \times  10^{-11}$ leads to a totally negligible $O(10^{-19})$ uncertainty in $a_e^{\mysmall \rm QED}$.


Almost one thousand four-loop diagrams contribute to the mass-independent coefficient $A_1^{(8)}$, and only few of them are known analytically. However, in a formidable effort that has its origins in the 1960s, Kinoshita and his collaborators calculated $A_1^{(8)}$ numerically~\cite{A_1^8}. Their latest result is
$
       A_1^{(8)}  =  -1.9106 \, (20),
$
from Ref.~\cite{Aoyama:2012wj}. In the same, very recent, article they also computed the tiny mass-dependent four-loop coefficients
$
     A_2^{(8)}(m_e/m_{\mu})  =  9.222 \,(66)  \times 10^{-4},
$
$
     A_2^{(8)}(m_e/m_{\tau}) = 8.24 \,(12)   \times 10^{-6},
$
and
$
     A_3^{(8)}(m_e/m_{\mu},m_e/m_{\tau}) = 7.465 \,(18) \times 10^{-7}.
$
Adding up these values one gets
\be
    C_4    =   -1.9097 \, (20).
\label{eq:C4}
\ee
The error $\delta C_4 = 0.0020$, caused by the numerical procedure, leads to an uncertainty of 
$5.8 \times 10^{-14}$ 
in $a_e^{\mysmall \rm QED}$. Independent work on the computation of the four-loop coefficient
is in progress~\cite{Laporta:2008zz}.


Very recently, Kinoshita and his collaborators completed the heroic calculation of the 12672 five-loop diagrams contributing to the tenth-order coefficient $A_1^{(10)}$~\cite{Aoyama:2012wj,A_1^10}. Their result is 
$
     A_1^{(10)}  =  9.16 \, (58).
$
They also computed the tiny mass-dependent term 
$
     A_2^{(10)}(m_e/m_{\mu})  =  -0.003 \,  82 \, (39).
$
As the tenth-order contribution of the $\tau$ lepton loops can be safely neglected, the sum of these two numbers provides the five-loop coefficient 
\be
    C_5    =   9.16 \, (58).
\label{eq:C5}
\ee
The error $\delta C_5 = 0.58$ leads to an uncertainty of $3.9 \times 10^{-14}$ in $a_e^{\mysmall \rm QED}$.

The complete {\small QED} contribution is given by
\be
a_{e}^{\mysmall \rm QED}(\alpha) =  \sum_{i=1}^\infty C_i (\alpha/\pi)^i.
\ee
As $(\alpha/\pi)^6 = 1.6 \times 10^{-16}$, terms of order $i \geq 6$ are assumed to be negligible at present.

\subsubsection{Electroweak and hadronic contributions}
\label{sec:ewhad}

The electroweak contribution is~\cite{Mohr:2012tt,EW}
\be
     a_{e}^{\mysmall \rm EW} = 0.2973 \, (52) \times 10^{-13}.
\label{eq:aEEW}
\ee
This precise value includes the two-loop contributions first computed in~\cite{EW}. 

The hadronic term,
\be
     a_{e}^{\mysmall \rm HAD} = 16.82 \, (16)\times 10^{-13},
\ee
is six times larger than the present experimental uncertainty $\delta a_{e}^{\mysmall \rm EXP}$, see \eq{eq:aEEX}.  
It is the sum of the following three contributions: the leading-order one,
$18.66 \, (11) \times 10^{-13}$,
very recently updated in~\cite{Nomura:2012sb} (see also~\cite{Jegerlehner:2009ry}),
the higher-order vacuum-polarization part, 
$-2.234 \, (14) \times 10^{-13}$~\cite{Nomura:2012sb} (see also~\cite{Jegerlehner:2009ry,Krause:1996rf}),
and the hadronic light-by-light term, 
$0.39 \, (13) \times 10^{-13}$~\cite{Jegerlehner:2009ry} (see also~\cite{Prades:2009tw}).
%

\subsubsection{Standard Model prediction of $a_e$ and value of $\alpha$}
\label{sec:smalpha}

The sum of the {\small QED} contribution $a_{e}^{\mysmall \rm QED}(\alpha)$ plus the hadronic
and weak terms discussed above yields the {\small SM} prediction of the electron $g$$-$$2$:
\be
	a_{e}^{\mysmall \rm SM}(\alpha) = a_{e}^{\mysmall \rm QED}(\alpha) +
	a_{e}^{\mysmall \rm EW}  + a_{e}^{\mysmall \rm HAD}
\ee
(the dependence on $\alpha$ of any contribution other than $a_{e}^{\mysmall \rm QED}$ is negligible). To compare it with experiment, we need the value of the fine-structure constant $\alpha$.  The latest determination of $\alpha$ by {\small CODATA}~\cite{Mohr:2012tt}, 
 \be
	\alpha \, \mbox{\footnotesize (CODATA)} = 1/137.035 \, 999 \, 074 \,(44)\,[0.32\,\mbox{ppb}],  
\label{eq:alphaCODATA10}
 \ee
cannot be employed for our purpose, as it is mainly driven by the value obtained equating the theoretical {\small SM} prediction of the electron $g$$-$$2$ with its measured value,
\be
      a_{e}^{\mysmall \rm SM}(\alpha) = a_{e}^{\mysmall \rm EXP}
\label{eq:SMvsEXP}
\ee
(thus assuming the absence of any significant new-physics contribution).\footnote{The identical value presently reported by the Particle Data Group (PDG)~\cite{PDG}, adopted from CODATA, cannot be used either.}
Indeed, solving \eq{eq:SMvsEXP} with the experimental value of \eq{eq:aEEX} we obtain
\be
	\alpha \, \mbox{($g$$-$$2$)} = 1/137.035 \, 999 \, 173 \, (34)\,[0.25\,\mbox{ppb}],
\label{eq:alphagm2}
\ee
in agreement with Ref.~\cite{Aoyama:2012wj}. This is the most precise value of $\alpha$ available today. The difference between this number and $\alpha \mbox{\footnotesize (CODATA)}$ in \eq{eq:alphaCODATA10} is mainly due to the very recent {\small QED} five-loop result of Ref.~\cite{Aoyama:2012wj}, which is included in our derivation, but not in the {\small CODATA}  one.

Clearly, in order to compute $a_{e}^{\mysmall \rm SM}(\alpha)$ and compare it with $a_{e}^{\mysmall \rm EXP}$ we must use a determination of $\alpha$ independent of the electron \gmt. At present, the
two most accurate ones are
\bea
\alpha \, (^{133}\mathrm{Cs})  &\!\!\!=& 
\!\!\!  1/137.036 \, 000 \, 0 \, (11) \,[7.7\,\mbox{ppb}],\label{eq:alphaCs}
\\
\alpha \, (^{87}\mathrm{Rb}) &\!\!\!=& 
\!\!\! 1/137.035 \, 999 \, 049 \, (90)\,[0.66\,\mbox{ppb}]. \label{eq:alphaRb10}
\eea
They differ by less than one standard deviation. The first value was obtained from the ratio $h/M_{\rm Cs}$ ($h$ is Planck's constant and $M_{\rm Cs}$ is the mass of the $^{133}\mathrm{Cs}$ atom), which was determined by measuring the atomic recoil frequency shift of photons absorbed or emitted by  $^{133}\mathrm{Cs}$ atoms using atom interferometry~\cite{Gerginov:2006zz}. The second was deduced from the measurement of the ratio $h/M_{\rm Rb}$ ($M_{\rm Rb}$ is the mass of the $^{87}\mathrm{Rb}$ atom) with an experimental scheme that combines atom interferometry with Bloch oscillation~\cite{Clade:2006zz,Cadoret2008,Bouchendira:2010es}. The values of $\alpha$ in eqs.~(\ref{eq:alphaCs},\ref{eq:alphaRb10}) were inferred from the ratios $h/M_{\rm Cs,Rb}$ combining them with the very precisely known Rydberg constant and the mass ratios $M_{\rm Cs,Rb} / m_e$~\cite{Mohr:2012tt}.
Given the higher precision of $\alpha(^{87}\mathrm{Rb})$ vs.\ $\alpha(^{133}\mathrm{Cs})$ (by more than one order of magnitude), the former is the value of $\alpha$ we employ to compute $a_{e}^{\mysmall \rm SM}(\alpha)$.
We note that $\alpha(^{87}\mathrm{Rb})$ agrees with $\alpha \mbox{($g$$-$$2$)}$ in \eq{eq:alphagm2} (the difference is 1.3 standard deviations), and its uncertainty $\delta\alpha(^{87}\mathrm{Rb})$ is larger than $\delta\alpha \mbox{($g$$-$$2$)}$ just by a factor of 2.7.

The {\small SM} prediction $a_{e}^{\mysmall \rm SM}(\alpha)$, computed with the fine-structure constant value $\alpha (^{87}\mathrm{Rb})$ of \eq{eq:alphaRb10}, is
\be
    a_e^{\scriptscriptstyle \rm SM} \!=\!
   115 \, 965 \, 218 \, 17.8 \, (0.6) (0.4) (0.2) (7.6) \times 10^{-13}.
\label{eq:aESM}
\ee
The first (second) error is determined by the uncertainty of the four(five)-loop {\small QED} coefficient, 
the third one is $\delta a_{e}^{\mysmall \rm HAD}$, 
and the last ($7.60 \times 10^{-13}$) is caused by the error $\delta\alpha(^{87}\mathrm{Rb})$. The uncertainties of the {\small EW} and two/three-loop {\small QED} contributions are totally negligible. When combined in quadrature, all these uncertainties yield 
$\delta a_e^{\scriptscriptstyle \rm SM} = 7.64 \times 10^{-13}$. 
Note that the present precision of the {\small SM} prediction, which is about three times worse than the experimental one, is limited by the uncertainty of the fine-structure constant 
$\alpha(^{87}\mathrm{Rb})$.

\subsubsection{Standard Model vs.\  measurement}
\label{sec:smm}

The {\small SM} value in \eq{eq:aESM} is in good agreement with the experimental one in \eq{eq:aEEX}. They differ by
\be
        \Delta a_e =      a_e^{\scriptscriptstyle \rm EXP} -
                                  a_e^{\scriptscriptstyle \rm SM} = 
                                  -10.5 \, (8.1) \times 10^{-13},
\label{eq:diff}
\ee 
{\it i.e.}\ 1.3 standard deviations, thus providing a beautiful test of {\small QED} at four-loop level! (The four-loop contribution to $a_{e}^{\mysmall \rm QED}$ is $-5.56 \times 10^{-11}$.) Once again, the uncertainty $\delta \Delta a_e = 8.1 \times 10^{-13}$ is
dominated by that of the {\small SM} prediction, through the error caused by $\delta\alpha(^{87}\mathrm{Rb})$.

\section{New Physics tests with $a_{e}$}
\label{sec:3}

New physics effects in the electron \gmt are usually expected to be comparable
with the {\small EW} contribution,
$     a_{e}^{\mysmall \rm EW} = 0.2973 \, (52) \times 10^{-13}$,
see \eq{eq:aEEW}, and therefore much smaller than the uncertainty $\delta \Delta a_e = 8.1 \times 10^{-13}$ reported above. Indeed, as we mentioned earlier, the commonly used {\small CODATA} (and {\small PDG}) value of $\alpha$ is mainly derived from $a_e^{\scriptscriptstyle \rm EXP}$ under this assumption. However, as we will discuss in the next section, the uncertainty
in $\Delta a_e$ is expected to be reduced. Then, the anomalous magnetic moment of the electron will provide us with important information on new physics effects.

\subsection{Future improvements in the determination of $\Delta a_e$}
 
As we showed in sections~\ref{sec:smalpha} and \ref{sec:smm}, the uncertainty $\delta \Delta a_e = 8.1 \times 10^{-13}$ is the result of the combination, in quadrature, of the following errors, in units of $10^{-13}$: 
\be
\underbrace{	(0.6)_{\rm QED 4} ,~~
			(0.4)_{\rm QED 5},~~ \phantom{\frac{1}{1}}
			(0.2)_{\rm HAD} }_{\displaystyle (0.7)_{\rm TH}^{\phantom{1}}},~~
	(7.6)_{\rm \delta \alpha},~~
	(2.8)_{\delta a_e^{\rm EXP}}.
\ee
The first one, $0.6 \times 10^{-13}$, is caused by the numerical procedure used to determine
the four-loop {\small QED} coefficient. It is already rather small and can be further reduced
to $0.1 \times 10^{-13}$ with a large scale numerical recalculation~\cite{KinoshitaLDMbook}.
Also the second error, induced by the five-loop QED term recently computed by Kinoshita and
his collaborators, is already small, and may soon drop to less than $0.1 \times 10^{-13}$~\cite{KinoshitaLDMbook}.
As the tiny -- but hard to reduce -- hadronic uncertainty is $0.16 \times 10^{-13}$ and those of the {\small EW} and two/three-loop {\small QED} contributions are totally negligible, the overall purely theoretical error of $\Delta a_e$ (i.e.\ the value of $\delta \Delta a_e$ obtained setting to zero $\delta a_e^{\rm \mysmall EXP}$ and the the error induced by $\delta \alpha$) is,
at present, $0.7 \times 10^{-13}$. This is likely to decrease even further, by a factor of two or three, in
a relatively near future.

Thanks to these recent theoretical improvements, the precision on $\Delta a_e$ is now limited only by the experimental uncertainties $\delta a_e^{\mysmall \rm EXP}$ and $\delta \alpha$. At present they affect $\delta \Delta a_e$ by $2.8 \times 10^{-13}$ ($\delta a_e^{\mysmall \rm EXP}$) and $7.6 \times 10^{-13}$ ($\delta \alpha$). It seems reasonable to expect a reduction of the
former error to a part in  $10^{-13}$ (or better) in ongoing efforts to improve the measurement of the electron (and positron) anomalous magnetic moment~\cite{GabrielseLDMBook,GabrielsePrivate}. Work is also in progress for a significant reduction of the latter error~\cite{Bouchendira:2010es,NezPrivate}.

In conclusion, a determination of $\Delta a_e$ at the level of $10^{-13}$ (or below) is a goal that can be achieved not too
far in the future with ongoing experimental work. As we will discuss in the next section, this will bring $a_e$ to play
a pivotal role in probing new physics in the leptonic sector.

\subsection{General structure of new-physics contributions}
\label{sec:model_independent}

The one-loop {\small SM} electroweak contribution to the anomalous magnetic
moment of the lepton $\ell$ ($\ell =e,\mu,\tau$) is
\beq
\left(a_\ell^{\mysmall \rm EW}\right)_{\mysmall \rm 1 loop} =
\frac{m_\ell^2}{(4\pi v)^2}\left( 1-\frac{4}{3}\sin^2\theta_{\rm W} + \frac{8}{3}
\sin^4\theta_{\rm W} \right)
\approx
2 \times 10^{-9}~  \frac{m^2_\ell}{m^2_\mu}\,,
\eeq
where $v=174~$GeV.
This can be viewed as a benchmark for contributions from new physics at the electroweak
scale. For the muon, this is about the same size as the observed discrepancy, see \eq{eq:gmu}.

New physics effects for the leptonic \gmt can be accounted for by means of the effective Lagrangian
\be
\mathcal L = e\frac{m_{\ell}}{2}
\left({\bar\ell}_{R}\sigma_{\mu\nu} A_{\ell\ell^{\prime}} \ell^{\prime}_{L} +
{\bar\ell^{\prime}}_{L}\sigma_{\mu\nu} A^{\star}_{\ell\ell^{\prime}} \ell_{R}\right)
F^{\mu\nu}\,\qquad \ell,\ell^{\prime} = e,\mu,\tau\,,
\label{eq:eff_lagr_LFV}
\ee
which describes dipole transitions ($\ell\to\ell^{\prime}\gamma$) in the leptonic sector.
Starting from eq.~(\ref{eq:eff_lagr_LFV}), we can evaluate $\Delta a_{\ell}$ as
\be
\Delta a_{\ell} = 2 m^{2}_{\ell}~{\rm Re}(A_{\ell\ell})\,.
\label{eq:gm2}
\ee
Let us consider now some new particles with typical mass $\Lambda_{\mysmall \rm NP}$ and
couplings $g_{\ell}^{L}$ and $g_{\ell}^{R}$ to left- and right-handed leptons $\ell$,
respectively. The one-loop new-physics contribution to the amplitude $A_{\ell\ell^{\prime}}$
is then of the form
\begin{equation}
A_{\ell\ell^{\prime}} =
\frac{1}{(4\pi\, \Lambda_{\mysmall \rm NP})^2}
\left[
\left(
g^{L}_{\ell k}\,g^{L*}_{\ell^{\prime} k} + g^{R}_{\ell k}\,g^{R*}_{\ell^{\prime} k}
\right) f_{1}(x_k) +
\frac{v}{m_{\ell}} \left(g^{L}_{\ell k}\,g^{R*}_{\ell^{\prime}k}\right) f_{2}(x_k)
\right]\,,
\label{eq:gm2_general}
\end{equation}
and therefore $\Delta a_{\ell}$ reads now
\be
\Delta a_{\ell} =
\frac{2 m^2_\ell}{(4\pi\, \Lambda_{\mysmall \rm NP})^2}
\left[
\left(|g^{L}_{\ell k}|^2 + |g^{R}_{\ell k}|^2\right) f_{1}(x_k)+
\frac{v}{m_\ell}{\rm Re}\left(g^{L}_{\ell k} g^{R*}_{\ell k}\right)f_{2}(x_k)
\right]\,.
\label{eq:dipoles_final1}
\ee
With $f_{1,2}$ we indicate loop functions which depend on ratios $(x_k)$ of unknown
masses of the new particles contributing to the amplitude $\ell\to\ell^{\prime}\gamma$,
and $k$ is a lepton flavor index.
In the term proportional to $f_1$, the chiral flip required by the dipole transition
occurs through a mass insertion in the external lepton line. In the term proportional
to $f_2$, the mass insertion is in the internal line of some new particle, thus 
explaining the parametric factor $v/m_\ell$. Although $f_2$ must be proportional to
the lepton Yukawa coupling, as a consequence of chiral symmetry, in practice this
term can become very sizeable whenever a new large coupling leads to a chiral enhancement.

In a broad class of theories beyond the {\small SM}, $g^{L,R}_\ell$ and $f_1$ are flavor
universal ({\it i.e.} are the same for any $\ell$) and $f_2$ vanishes, such that
\be
\frac{\Delta a_{\ell_i}}{\Delta a_{\ell_j}}
=\left( \frac{m_{\ell_i}}{m_{\ell_j}}\right)^2 \, .
\ee
We will refer to this case as ``naive scaling'' ({\small NS}).  {\small NS} applies, for instance,
if the new particles have an underlying SU(3) flavor symmetry in their mass spectrum and
in their couplings to leptons (which is the case for gauge interactions).

An interesting consequence of {\small NS} is that an explanation of the muon \gmt
anomaly makes definite predictions for new effects in the anomalous magnetic moments of
electron and $\tau$,
\bea
&&\Delta a_e = \left( \frac{\Delta a_\mu}{3\times 10^{-9}}\right) ~0.7\times 10^{-13}\,, \\
&&\Delta a_\tau = \left( \frac{\Delta a_\mu}{3\times 10^{-9}}\right) ~0.8\times 10^{-6}.
\eea
It is very exciting that the sensitivity in $\Delta a_e$ is not far from what is required
to test whether the discrepancy in the muon \gmt also manifests itself in the electron
$g$$-$$2$ under the {\small NS} hypothesis. Thus, determining $\Delta a_e$ with a
precision below $10^{-13}$ is an important goal with rich physics consequences.
The measurement of $a_{\tau}$ may play a similar role if a precision below $10^{-6}$ will be attained (see sect.~\ref{sec:tau_gmt}).  

Although the {\small NS} case is especially simple and common to a large class
of new-physics interactions, it is by no means the only possibility. Many theories make
predictions beyond {\small NS}, either because of the chirally-enhanced term proportional
to $f_2$, or because of lepton non-universality in the couplings $(g^{L,R}_\ell)$ or the mass spectrum ($f_1$). For instance, in a multi-Higgs doublet model, the couplings $g^{L,R}_\ell$
are related to Yukawa couplings and therefore one would find the scaling
$\Delta a_{\ell_i}/\Delta a_{\ell_j} = m^{4}_{\ell_i}/m^{4}_{\ell_j}$. A different example
is offered by supersymmetry with non-degenerate sleptons, in which $\Delta a_{\ell}$ loses
its correlation with lepton masses.

The case of non-naive scaling is interesting from the theoretical point of view because
it allows for exploration of the structure of lepton symmetries. It is also interesting
experimentally because it can lead to effects in $\Delta a_e$ testable already with
present sensitivity. Moreover, as we will show with various examples in the next sections,
any lepton non-universality in $\Delta a_\ell$ can be related to other experimental
observables, offering the possibility of cross-checking new physics effects.

The underlying $\ell\to\ell^{\prime}\gamma$ transition described by the effective
Lagrangian of eq.~(\ref{eq:eff_lagr_LFV}) can generate, in addition to the anomalous
magnetic moments $\Delta a_{\ell}$, also lepton flavor violating ({\small LFV}) processes ($\ell \neq \ell^{\prime}$),
such as $\mu\to e\gamma$, and CP violating effects, such as the leptonic electric 
dipole moments ({\small EDM}s, $d_\ell$),
\be
\frac{{\rm BR}(\ell\to \ell^{\prime}\gamma)}{{\rm BR}(\ell\to \ell^{\prime}\nu_\ell\bar\nu_{\ell^\prime})} =
\frac{48\pi^3\alpha}{G^2_F}
\left( |A_{\ell\ell^{\prime}}|^2 + |A_{\ell^{\prime}\ell}|^2 \right)\,,\qquad
\frac{d_{\ell}}{e} = m_{\ell}~{\rm Im}(A_{\ell\ell})\,.
\label{eq:BR_LFV}
\ee
%
%
%
Using the general parametrization for $A_{\ell\ell^{\prime}}$ of
eq.~(\ref{eq:gm2_general}) we find
\bea
\frac{{\rm BR}(\ell\to \ell^{\prime}\gamma)}{{\rm BR}(\ell\to \ell^{\prime}\nu_\ell\bar\nu_{\ell^\prime})} &\!\!\!\!\!=\!\!\!\!\!&
\frac{48\pi\alpha G^{-2}_F}{(4\pi\Lambda_{\mysmall \rm NP})^4}
\left(\left| \left(g^{L}_{\ell k}~g^{L*}_{\ell^{\prime} k} + g^{R}_{\ell k}~ g^{R*}_{\ell^{\prime} k}
\right) f_{1}(x_k) + \frac{v}{m_{\ell}} g^{L}_{\ell k}~g^{R*}_{\ell^{\prime} k}
f_{2}(x_k) \right|^2 +
\ell \leftrightarrow \ell^{\prime}\right)\,,
\nonumber\\
\frac{d_{\ell}}{e} &=& \frac{v}{(4\pi\, \Lambda_{\mysmall \rm NP})^2} \,\,
{\rm Im}\left(g^{L}_{\ell k}~g^{R*}_{\ell k}\right)f_{2}(x_k)\,.
\label{eq:dipoles_final}
\eea
On general grounds, one would expect that, in concrete {\small NP} scenarios, $\Delta a_{\ell}$, 
$d_{\ell}$ and ${\rm BR}(\ell\to \ell^{\prime}\gamma)$, are correlated. In practice,
their correlations depend on the unknown flavor and CP structure of the couplings $g^{L}$
and $g^{R}$, and thus we cannot draw any firm conclusion. In the following, we will point
out the general conditions that have to be fulfilled by any {\small NP} theory in order to account
for large effects in $\Delta a_{\ell}$ while satisfying the constraints from $d_{\ell}$
and ${\rm BR}(\ell\to \ell^{\prime}\gamma)$.

Regarding the leptonic {\small EDM}s, we find the following {\it model-independent} relations
\bea
d_{e} &\simeq&
\left(\frac{\Delta a_{e}}{7 \times 10^{-14}}\!\right)
10^{-24}~\tan\phi_{e}~~ e~{\rm cm}\,,
\nonumber\\
d_{\mu} &\simeq&
\left(\frac{\Delta a_{\mu}}{3 \times 10^{-9}}\right)
2\times 10^{-22}~\tan\phi_{\mu}~~ e~{\rm cm}\,,
\nonumber\\
d_{\tau} &\simeq&
\left(\frac{\Delta a_{\tau}}{8 \times 10^{-7}}\right)
4\times 10^{-21}~\tan\phi_{\tau}~~ e~{\rm cm}\,,
\label{eq:gm2_vs_edm}
\eea
where we have defined $\phi_{\ell}={\rm arg}(A_{\ell\ell})$. We have normalized $\Delta a_{\mu}$
to the central value of the current anomaly and  $\Delta a_{e,\tau}$ to their values in naive scaling, $\Delta a_{\ell}/\Delta a_{\mu}=m^2_\ell/m^2_\mu$.

From eq.~(\ref{eq:gm2_vs_edm}) we learn that an explanation of the $\Delta a_{\mu}$
anomaly implies, for a natural CPV phase $\phi_{\mu}\sim \mathcal{O}(1)$,
a {\it model-independent} upper bound on $d_{\mu} \lesssim 3\times 10^{-22} \, e~$cm
which is still far from the current bound $d_{\mu} \lesssim 10^{-18} \, e~$cm,
but well within the expected future sensitivity $d_{\mu} < 10^{-24} \, e~$cm~\cite{Semertzidis:2011zz}.
Therefore, any experimental effort to improve the resolution on $d_{\mu}$ would be valuable.

On the other hand, the electron {\small EDM} 
imposes a bound on the corresponding CPV phase $\phi_e$
at the level of $10^{-3}$, if {\small NS} is at work. Such a condition could be realized
for instance if $\phi_e$ is generated radiatively while $\phi_{\mu}$ arises already at the
tree level. Going beyond {\small NS}, one could also envisage scenarios where the 
electronic dipoles are suppressed compared to the muonic dipoles because of hierarchical couplings $g^{L,R}_e \ll g^{L,R}_\mu$, as it happens for instance in a multi-Higgs doublet
model where $g^{L,R}_\ell$ are related to Yukawa couplings. In general, as shown by eq.~(\ref{eq:dipoles_final}), the {\small EDM}s (but not $\Delta a_{\mu}$) vanish if $g^{L} = g^{R}$ as it could arise in a left-right symmetric theory.

Now we discuss the potential constraints on $\Delta a_\ell$ arising from
{\small LFV} processes. An inspection of eq.~(\ref{eq:dipoles_final}) shows
that ${\rm BR}(\ell\to \ell^\prime\gamma)$ and $\Delta a_{\ell}$ are generally
correlated as ${\rm BR}(\ell\to \ell^\prime\gamma)\sim(\Delta a_{\mu})^2~|\theta_{\ell\ell^\prime}|^2$ where $\theta_{\ell\ell^\prime}$
stands for the relevant flavor mixing angle, which is approximately $\theta_{\ell\ell^\prime}\sim\frac{g_{\ell k}g^{*}_{\ell^\prime k}}{g_{\ell k} g^{*}_{\ell k}}$,
see eq.~(\ref{eq:dipoles_final}). In particular we have
\bea
{\rm BR}(\mu\to e\gamma) &\approx& 2 \times 10^{-12}
\left(\frac{\Delta a_{\mu}}{3 \times 10^{-9}}\right)^2
\left(\frac{\theta_{e\mu}}{3 \times 10^{-5}}\right)^2\,,
\nonumber\\
{\rm BR}(\tau\to \ell\gamma) &\approx& 4\times 10^{-8}
\left(\frac{\Delta a_{\mu}}{3 \times 10^{-9}}\right)^2
\left(\frac{\theta_{\ell\tau}}{ 10^{-2}}\right)^2\,.
\label{eq:gm2_vs_lfv}
\eea
{\small LFV} contributions to $\Delta a_\ell$ can be particularly important as they are typically
chirally enhanced by $m_\tau/m_\ell$, as we will discuss in the case of supersymmetry.
Such a chiral enhancement is not effective in $\tau$ {\small LFV} processes and therefore, in
this case, it might be possible to keep ${\rm BR}(\tau\to \ell\gamma)$ under control,
while generating large effects especially for $\Delta a_e$.


Before turning our attention to some prototype theories beyond the {\small SM} that predict
non-naive scaling in $\Delta a_\ell$, we will now briefly summarize the status of the anomalous magnetic moment of the $\tau$.

\subsection{Status of the $\tau$ anomalous magnetic moment}
\label{sec:tau_gmt}

The present experimental sensitivity of the $\tau$ lepton $g$$-$$2$ is only $10^{-2}$. In fact, while the {\small SM} prediction for $a_{\tau}$ is precisely known~\cite{Eidelman:2007sb},
\be
a_{\tau}^{\mysmall \rm SM} = 117 \, 721 \, (5) \times 10^{-8},
\label{eq:atSM}
\ee
the very short lifetime of this lepton ($2.9 \times 10^{-13}$ s) makes it very difficult to determine its anomalous magnetic moment by measuring its spin precession in a magnetic field, like in the electron and muon $g$$-$$2$ experiments. Instead, experiments focused on high-precision measurements of $\tau$ pair production in various high-energy processes and comparison of the measured cross sections with the {\small SM} predictions.

The present {\small PDG} limit on the $\tau$  $g$$-$$2$ was derived by the {\small DELPHI} collaboration from $e^+e^- \to e^+e^-\tau^+\tau^-$ total cross section measurements at {\small LEP2}:
$
                         -0.052 < a_{\tau}^{\mysmall \rm EXP} < 0.013
$
at $95\%$ confidence level~\cite{Abdallah:2003xd}. This reference also quotes the result in the form:
\be
                         a_{\tau}^{\mysmall \rm EXP} = -0.018 (17).
\label{eq:atEXP}
\ee
Comparing eqs.~(\ref{eq:atSM}) and (\ref{eq:atEXP}) (their difference is roughly one standard deviation), it is clear that the sensitivity of the best existing measurements is still more than an order of magnitude worse than needed. In \cite{GonzalezSprinberg:2000mk}, the reanalysis of various measurements of the cross section of the process $e^+e^- \to \tau^+\tau^-$, the transverse $\tau$ polarization and asymmetry at {\small LEP} and {\small SLD}, as well as of the decay width $\Gamma(W \to \tau\nu_{\tau})$ at {\small LEP} and Tevatron allowed the authors to set a stronger model-independent limit on new physics contributions:
$
                       -0.007 < a_{\tau}^{\mysmall \rm NP}  < 0.005.
$

The possibility to improve these limits is certainly not excluded. Future high-luminosity $B$ factories, such as Belle-II~\cite{Aushev:2010bq} or SuperB~\cite{Lusiani:2011zz}, offer new opportunities to improve the determination of the $\tau$ magnetic properties. The authors of \cite{Bernabeu:2007-8} proposed to measure the $\tau$ anomalous magnetic moment form factor at the $\Upsilon$ resonance with sensitivities down to $10^{-5}$ or $10^{-6}$, and work is in progress for the determination of $a_{\tau}$ via the measurement of radiative leptonic $\tau$ decays~\cite{RadLepTauDec}.

\section{Supersymmetry and $a_{e}$}
\label{sec:susy}

The supersymmetric contribution to $a_\ell$ comes from loops with exchange of chargino/sneutrino or neutralino/charged lepton. Therefore, violations of ``naive scaling'' can arise through sources of non-universalities in the slepton mass matrices. Such non-universalities can be realized in two ways.
\begin{enumerate}
\item Lepton flavor conserving ({\small LFC}) case. The charged slepton mass matrix violates the global non-abelian flavor symmetry, but preserves U(1)$^3$. This case is characterized by non-degenerate sleptons ($m_{\tilde e}\neq m_{\tilde\mu}\neq m_{\tilde\tau}$) but vanishing mixing angles because
of an exact alignment, which ensures that Yukawa couplings and the slepton mass matrix can be simultaneously diagonalized in the same basis.
\item Lepton flavor violating ({\small LFV}) case. The slepton mass matrix fully breaks flavor symmetry
up to U(1) lepton number, generating mixing angles that allow for flavor transitions.
Lepton flavor violating processes, such as $\mu \to e \gamma$, provide stringent constraints
on this case. However, because of flavor transitions, $a_e$ and $a_\mu$ can receive new large contributions proportional to $m_\tau$ (from a chiral flip in the internal line of
the loop diagram), giving a new source of non-naive scaling.
\end{enumerate}
We will discuss these two cases separately.

\subsection{Lepton flavor conserving case}

In the {\small LFC} case, we assume non-degenerate slepton masses for different families
($m_{\tilde e}\neq m_{\tilde\mu}\neq m_{\tilde\tau}$) but flavor alignment between
lepton and slepton mass matrices. This is reminiscent of the well-known alignment mechanism~\cite{Nir:1993mx}, which was proposed to solve the supersymmetric flavor
problem in the quark sector by aligning the down quark/squark mass matrices and which
might arise naturally in the context of abelian flavor models~\cite{Leurer:1993gy}.

In this case the supersymmetric contribution to $\Delta a_\ell^{\mysmall \rm LFC}$
is given by the following approximate expression
\bea
\label{eq:gm2_MI}
\Delta a_\ell^{\mysmall \rm LFC} &=&
\frac{\alpha m^{2}_{\ell}}{4\pi\sin^2\theta_W}
\frac{{\rm Re}(\mu M_{2} \tan\beta)}{m_{\tilde\ell}^{2}(M_{2}^2\!-\!\mu^2)}
~\bigg[ \frac{1}{2}f_{n}(x_{2},x_{\mu}) - f_{c}(x_{2},x_{\mu})\bigg]
\nonumber\\
&+&\frac{\alpha m^{2}_{\ell}}{8\pi\cos^2\theta_W}
\frac{{\rm Re}(\mu M_{1} \tan\beta)}{m_{\tilde\ell}^{2}(M_{1}^2\!-\!\mu^2)}
~f_{n}(x_{1},x_{\mu})
\nonumber\\
&-&\frac{\alpha m^{2}_{\ell}}{4\pi\cos^2\theta_W}
\frac{{\rm Re}(A_\ell M_{1} -\mu M_{1}\tan\beta)}{m_{\tilde\ell}^{2}}~
g_{n}(x_{1})\,\qquad \ell = e,\mu,\tau\,.
\eea
The loop functions $f_n$, $f_c$ and $g_n$ (with $x_{1(2)} = M_{1(2)}^2/m_{\tilde\ell}^2$,
and $x_{\mu}=\mu^2/m_{\tilde\ell}^2$) are given in the appendix, and we
use standard notation for the supersymmetric parameters.
%
%
%
%
In our numerical analysis we use the exact expressions for $\Delta a_\ell^{\mysmall \rm LFC}$
in the mass eigenstate basis~\cite{Moroi:1995yh}.

In the illustrative case of a single mass scale ($M_{1,2}=\mu=m_{\tilde\ell}$)
and if ${\rm arg}(\mu M_{1,2})= {\rm arg}(A_\ell M_{1}) = 0$, the result simplifies
to\footnote{The overall sign of the supersymmetric contribution to $a_\ell$
is a free parameter, since it is determined by ${\rm arg}(\mu M_{1,2})$, which
is invariant under field phase redefinition.}
\bea
\Delta a_\ell^{\mysmall \rm LFC} &=&
\frac{5\alpha_2}{48\pi}\,\frac{m^{2}_{\ell}}{m_{\tilde\ell}^{2}}\,\tan\beta +
\frac{\alpha_Y}{24\pi}\,\frac{m^{2}_{\ell}}{m_{\tilde\ell}^{3}} A_\ell
\nonumber\\
&\approx&
3 \times 10^{-9} \left(\frac{m_\ell}{m_\mu}\right)^2
\left(\frac{100~{\rm GeV}}{m_{\tilde\ell}}\right)^2
\left[\left(\frac{\tan\beta}{3}\right) + 
0.1 \left(\frac{A_\ell}{3 m_{\tilde\ell}}\right)\right]\,.
\label{eq:gm2_susy_1}
\eea
This shows that the discrepancy in the muon \gmt can be explained by
supersymmetric particles with masses around 100 (300)~GeV, for $\tan\beta =3~(20)$.
On the other hand, the contribution proportional to the term $A_\ell$,
which could induce in principle a violation of the {\small NS} if $A_e/A_\mu \neq m_e/m_\mu$,
is too small to explain the muon \gmt anomaly after imposing the vacuum stability
bound $|A_\ell|/m_{\tilde \ell} \lesssim 3$.

\begin{figure*}[p]
\centering
\includegraphics[width=0.49\textwidth]{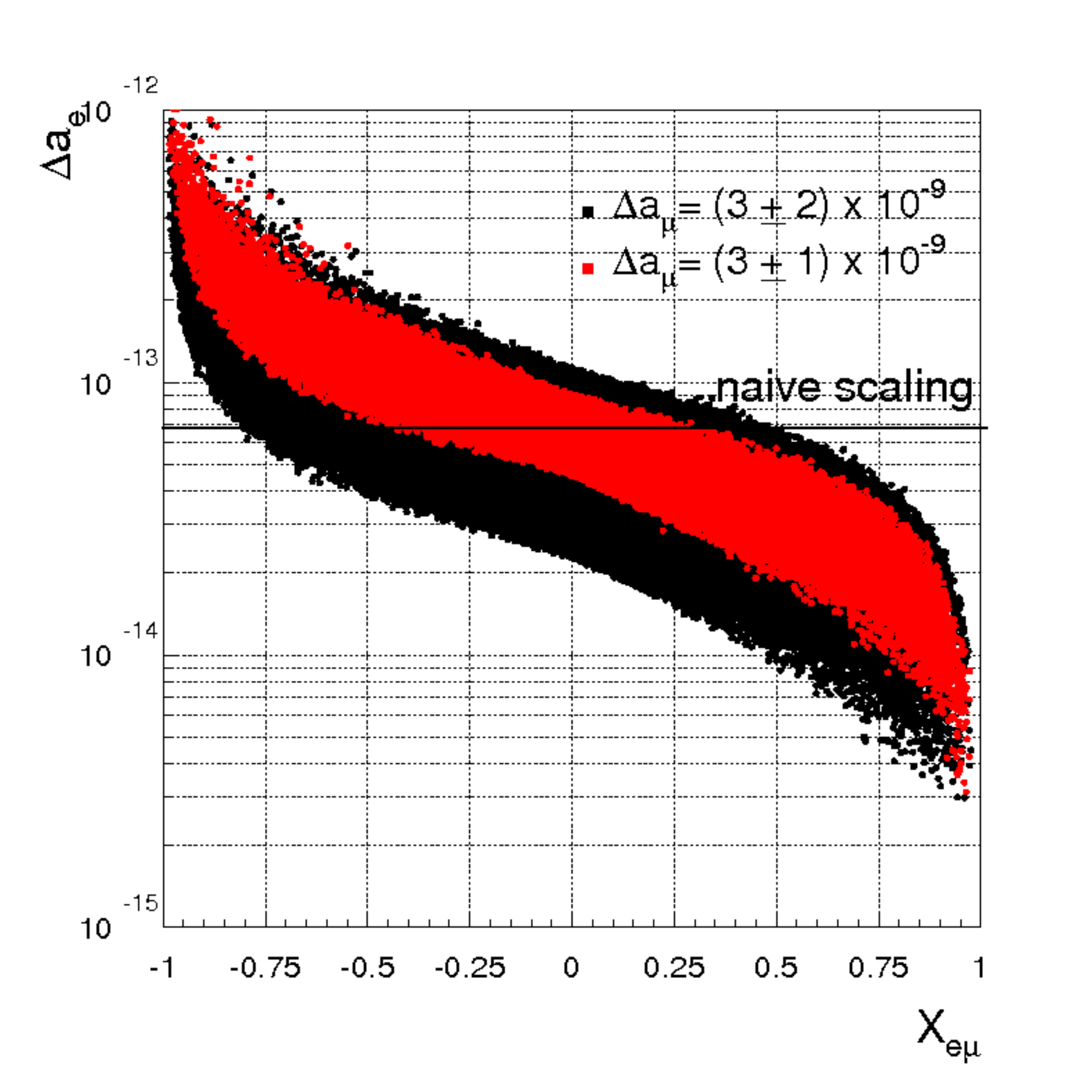}
\includegraphics[width=0.49\textwidth]{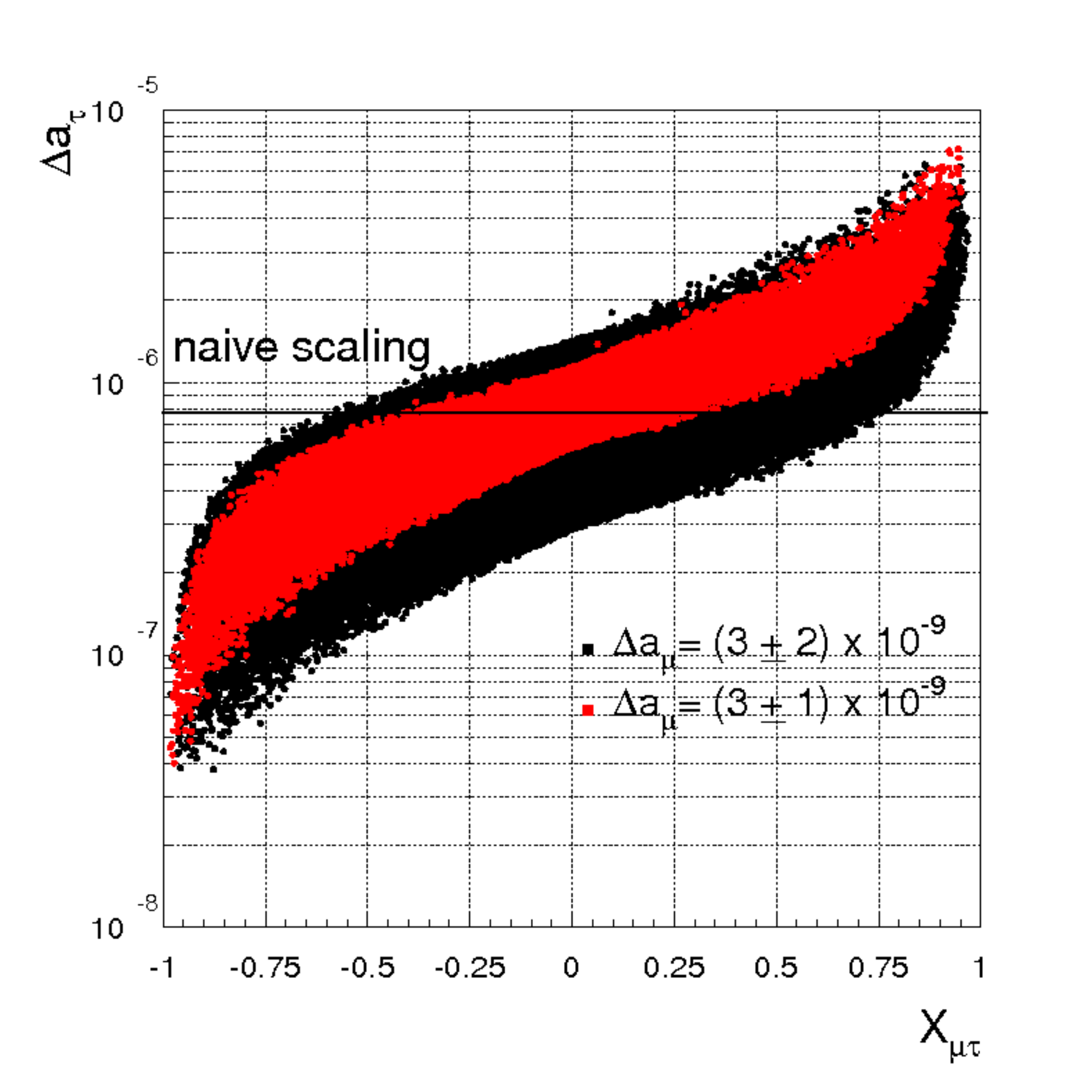}
\caption{
Left: $\Delta a_{e}$ as a function of $X_{e\mu}=(m^{2}_{\tilde e}-m^{2}_{\tilde \mu})/(m^{2}_{\tilde e}+m^{2}_{\tilde\mu})$.
Right: $\Delta a_{\tau}$ as a function of $X_{\mu\tau}=(m^{2}_{\tilde \mu}-m^{2}_{\tilde \tau})/(m^{2}_{\tilde \mu}+m^{2}_{\tilde\tau})$.
Black points satisfy the condition $1\leq \Delta a_{\mu}\times 10^{9}\leq 5$,
while red points correspond to $2\leq \Delta a_{\mu}\times 10^{9}\leq 4$.}
\label{fig:LFC}
\end{figure*}

Assuming that sleptons are the heaviest particles running in the loop, we find
\bea
&&\Delta a_e \approx \Delta a_\mu~\frac{m^{2}_{e}}{m^{2}_{\mu}}\frac{m^{2}_{\tilde\mu}}{m^{2}_{\tilde e}}
\approx
\frac{m^{2}_{\tilde\mu}}{m^{2}_{\tilde e}}\left( \frac{\Delta a_\mu}{3\times 10^{-9}}\right) 10^{-13}\,, \nonumber \\
&&\Delta a_\tau \approx \Delta a_\mu~\frac{m^{2}_{\tau}}{m^{2}_{\mu}}\frac{m^{2}_{\tilde\mu}}{m^{2}_{\tilde \tau}}
\approx
\frac{m^{2}_{\tilde\mu}}{m^{2}_{\tilde \tau}} \left( \frac{\Delta a_\mu}{3\times 10^{-9}}\right)  10^{-6} .
\label{eq:gm2_susy_2}
\eea
For values of $\Delta a_\mu$ explaining the muon $g$$-$$2$, non-degenerate sleptons at the
level $m_{\tilde\mu}\approx 3\, m_{\tilde e}$ lead to $\Delta a_e\approx 10^{-12}$,
which is at the limit of present experimental sensitivity. This naive expectation is
confirmed by a complete numerical analysis, as illustrated in the left panel of fig.~\ref{fig:LFC},
which shows $\Delta a_{e}$ as a function of the normalized selectron/smuon mass splitting 
$X_{e\mu}=(m^{2}_{\tilde e}-m^{2}_{\tilde \mu})/(m^{2}_{\tilde e}+m^{2}_{\tilde\mu})$.
The plot has been obtained by scanning the supersymmetric parameters that account for
the $(g-2)_\mu$ anomaly at the level of $1\leq \Delta a_{\mu}\times 10^{9}\leq 5$ (black points),
or $2\leq \Delta a_{\mu}\times 10^{9}\leq 4$ (red points in the inner region).
We have imposed all current bounds on sparticle masses arising from direct searches
and required $M_1 \leq 100$~GeV, $M_2 = 2 M_1$, $\mu \leq 200$~GeV, $3\leq\tan\beta\leq 30$, $m_{{\tilde e}, {\tilde \mu}}\leq 500$~GeV. As can be clearly seen, values of $\Delta a_{e}$
almost up to $10^{-12}$ are reachable for highly non-degenerate selectron and smuon
masses such that $|X_{e\mu}|\sim 1$.

In the right panel of fig.~\ref{fig:LFC} we show $\Delta a_{\tau}$ as a function of $X_{\mu\tau}=(m^{2}_{\tilde \mu}-m^{2}_{\tilde \tau})/(m^{2}_{\tilde \mu}+m^{2}_{\tilde\tau})$. The plot has been obtained following the same procedure described for the left panel. We observe that $\Delta a_{\tau}$ can reach the level of $10^{-5}$ if $m_{\tilde\tau}\ll m_{\tilde\mu}$, therefore violating {\small NS}, whose prediction is shown by the horizontal line
in fig.~\ref{fig:LFC}. The condition $m_{\tilde\tau}\ll m_{\tilde\mu}$ is justified
in the so-called split family models~\cite{io,altri}, where the first two generations 
of squarks and leptons are substantially heavier than the third generation.

Before concluding this section, it is worth to comment about the potential constraints
arising from the leptonic {\small EDM}s, on the same line done in the model-independent analysis
of sect.~\ref{sec:model_independent}. From eq.~(\ref{eq:gm2_vs_edm}) we learn that
$\Delta a_e \approx 10^{-13(12)}$ is allowed only if ${\rm arg}(\mu M_{1,2})\lesssim 10^{-3(4)}$.
Moreover, since we are dealing now with {\it flavor-blind} CPV phases, that is $\phi=\phi_{e}=\phi_{\mu}=\phi_{\tau} = {\rm arg}(\mu M_{1,2})$, the electron {\small EDM} limit
$d_{e}\lesssim 1.5\times 10^{-27}e~$cm implies the bounds $d_{\mu}\lesssim 3\times 10^{-25}e~$cm
and $d_{\tau}\lesssim 5\times 10^{-24}e~$cm, if
$m_{\tilde e}\simeq m_{\tilde \mu} \simeq m_{\tilde \tau}$.
Yet, values of $d_\mu$ as high as the model-independent upper bound $d_{\mu} \lesssim  10^{-22}e~{\rm cm}$
of eq.~(\ref{eq:gm2_vs_edm}) could be reached if $m_{\tilde e}\gg m_{\tilde \mu}$, in which case the supersymmetric contributions to $\Delta a_{e}$ and $d_e$ could be made negligible.

\subsubsection{Correlation between $a_{e}$ and violation of lepton universality in LFC}
\label{sec:naive_scaling}

In supersymmetric theories, {\small NS} violations for the leptonic
$(g-2)_\ell$ can arise through sources of non-universalities in the slepton masses.
In turn, these non-universalities will induce violations of lepton flavor universality
in low- and high-energy processes such as $P\to\ell\nu$, $\tau\to P\nu$ (where $P=\pi,K$),
$\ell_i\to \ell_j\bar\nu\nu$, $Z\to\ell\ell$ and $W\to\ell\nu$ through loop effects.
Lepton universality has been probed at the few per-mill level so far, see table~
\ref{table:LFU}. It is interesting to study the correlation between such violations
of lepton universality and departures from {\small NS} for $\Delta a_{\ell}$.
Taking for example the process $P\to\ell\nu$, we can define the quantity
\beq
\frac{(R_P^{e/\mu})_{\mysmall \rm EXP}}{(R_P^{e/\mu})_{\mysmall \rm SM}}
= 1 + \Delta r_P^{e/\mu} ~.
\label{RPemu}
\eeq
Here $(R_P^{e/\mu})_{\mysmall \rm SM}=\Gamma(P\to e\nu)_{\mysmall \rm SM}/\Gamma(P\to\mu\nu)_{\mysmall \rm SM}$ and $(R_P^{e/\mu})_{\mysmall \rm EXP}=\Gamma(P\to e\nu)_{\mysmall \rm EXP}/\Gamma(P\to \mu\nu)_{\mysmall \rm EXP}$ so
that $\Delta r_P^{e/\mu}\neq 0$ signals the presence of new physics violating
lepton universality.

\begin{table}
\centering
\renewcommand{\arraystretch}{1.5}
\begin{tabular}{lcc}
\hline\hline
Channel & $\Delta r^{e/\mu} $ \\
\hline
\hline
$\Gamma(\pi \rightarrow e \, \bar{\nu}_e)/\Gamma(\pi \rightarrow \mu \,\bar{\nu}_\mu)$ &
$- 0.0045 \pm 0.0032 $~\cite{TRI-PP-92-15} \\
$\Gamma(K \rightarrow e \,\bar{\nu}_e)/\Gamma(K \rightarrow \mu \,\bar{\nu}_\mu)$ &
$0.004 \pm  0.004 $~\cite{NA62:2011} \\
$\Gamma(K \rightarrow \pi \, e \,\bar{\nu}_e)/\Gamma(K \rightarrow \pi \, \mu \,\bar{\nu}_\mu)$ & 
$ -0.002 \pm 0.004 $~\cite{Kl3} \\
$\Gamma(Z \rightarrow e^+ e^-)/\Gamma(Z \rightarrow \mu^+ \mu^-)$ & $-0.0010 \pm 0.0026$
~\cite{PDG,LEPEWWG,LEPEWWG_SLD:06,Abbaneo:2001ix} \\
$\Gamma(W \rightarrow e \,\bar{\nu}_e)/\Gamma(W \rightarrow \mu \,\bar{\nu}_\mu )$ &
$ 0.017 \pm 0.019 $~\cite{PDG,LEPEWWG,LEPEWWG_SLD:06,Abbaneo:2001ix} \\
 $\Gamma(\tau \rightarrow \nu_\tau \,e \,\bar{\nu}_e)/\Gamma(\tau \rightarrow \nu_\tau\, \mu \,\bar{\nu}_\mu )$ & $ -0.0036 \pm 0.0028 $~\cite{BABAR-Univ:10} \\
\hline\hline
Channel & $\Delta r^{\mu /\tau} $ \\
\hline
\hline
$\Gamma(\pi \rightarrow \mu \,\bar{\nu}_\mu)/ \Gamma(\tau  \rightarrow \pi \, \nu_\tau)$ & $0.016\pm 0.008 $ \\
$\Gamma(K \rightarrow \mu \, \bar{\nu}_\mu)/\Gamma(\tau  \rightarrow K \, \nu_\tau)$ &
$0.037 \pm 0.016 $~\cite{BABAR-Univ:10} \\
$\Gamma(Z \rightarrow \mu^+ \mu^-)/\Gamma(Z \rightarrow \tau^+ \tau^-)$ &
$-0.0011 \pm 0.0034$~\cite{PDG,LEPEWWG,LEPEWWG_SLD:06,Abbaneo:2001ix} \\
$\Gamma(W \rightarrow \mu \, \bar{\nu}_\mu)/\Gamma(W \rightarrow \tau \, \bar{\nu}_\tau )$ &
$ -0.060 \pm 0.021 $~\cite{PDG,LEPEWWG,LEPEWWG_SLD:06,Abbaneo:2001ix} \\
 $\Gamma(\mu \rightarrow \nu_\mu\, e\, \bar{\nu}_e ) / \Gamma(\tau \rightarrow \nu_\tau\, e \,\bar{\nu}_e)$ & $ -0.0014 \pm 0.0044 $~\cite{BABAR-Univ:10} \\
\hline\hline
Channel & $\Delta r^{e /\tau} $ \\
\hline
\hline
$\Gamma(Z \rightarrow e^+ e^-)/\Gamma(Z \rightarrow \tau^+ \tau^-)$ &
$-0.0020 \pm 0.0030$~\cite{PDG,LEPEWWG,LEPEWWG_SLD:06,Abbaneo:2001ix} \\
$\Gamma(W \rightarrow e \, \bar{\nu}_e)/\Gamma(W \rightarrow \tau \, \bar{\nu}_\tau )$ &
$ -0.044 \pm 0.021 $~\cite{PDG,LEPEWWG,LEPEWWG_SLD:06,Abbaneo:2001ix} \\
$\Gamma(\mu \rightarrow \nu_\mu\, e\,\bar{\nu}_e ) /
\Gamma(\tau \rightarrow \nu_\tau\,\mu\,\bar{\nu}_\mu)$ & $ -0.0032 \pm 0.0042 $~\cite{BABAR-Univ:10} \\
\hline\hline
\end{tabular}
\renewcommand{\arraystretch}{1}
\caption{Experimental limits on $\Delta r^{\ell/\ell^\prime}$ with
$\ell,\ell^\prime = e, \mu, \tau$.
}
\label{table:LFU}
\end{table}

Within supersymmetry, in the absence of {\small LFV} sources, $\Delta r_P^{e/\mu}$ is
induced at the loop level by box, wave function renormalization and vertex
contributions from sparticle exchange. The complete calculation of $\mu$ decay
in supersymmetry~\cite{Chankowski:1993eu,Krawczyk:1987zj} can be easily applied
to meson decays and we have included the full result in our numerical analysis.
Neglecting box contributions, which are suppressed by heavy squark masses, the
parametrical structure of $\Delta r_{\rm P}^{e/\mu}$ is
\begin{equation}
\Delta r_{\rm P}^{e/\mu}
\sim
\frac{\alpha}{4\pi}
\left(\frac{m^{2}_{\tilde e} - m^{2}_{\tilde\mu}}{m^{2}_{\tilde e} + m^{2}_{\tilde\mu}}\right)
\frac{v^2}{{\rm min}(m^{2}_{\tilde{e},\tilde{\mu}})}\,,
\label{eq:LFU}
\end{equation}
where the term $v^2/{\rm min}(m^{2}_{\tilde{e},\tilde{\mu}})$ stems from
SU(2) breaking effects.

Within supersymmetry, such SU(2) breaking sources arise from left-right soft breaking
terms, from mixing terms in the chargino/neutralino mass matrices, or from D-terms.
For highly non degenerate sleptons
and if ${\rm min} (m^{2}_{\tilde{e},\tilde{\mu}})\sim v^2$, $\Delta r_{\rm P}^{e/\mu}$
can reach the few per-mill level.

Given that the {\small NP} sensitivity of $\pi\to\ell\nu$ and $K\to\ell\nu$ to the above effects
is the same, and that their present experimental resolutions are comparable, both $\pi\to\ell\nu$
and $K\to\ell\nu$ represent useful probes of this scenario.
Future experiments at {\small TRIUMF} and {\small PSI} aim at testing lepton universality in $\pi\to\ell\nu$
at the level of $< 1\times 10^{-3}$ and $5\times 10^{-4}$, respectively. Both the {\small NA62}
experiment at {\small CERN} and the {\small KLOE-2} experiment will continue improving their sensitivity
aiming at a test of lepton universality in $K \to \ell \nu$ at the few per-mill level.

\begin{figure*}[p]
\centering
\includegraphics[width=0.49\textwidth]{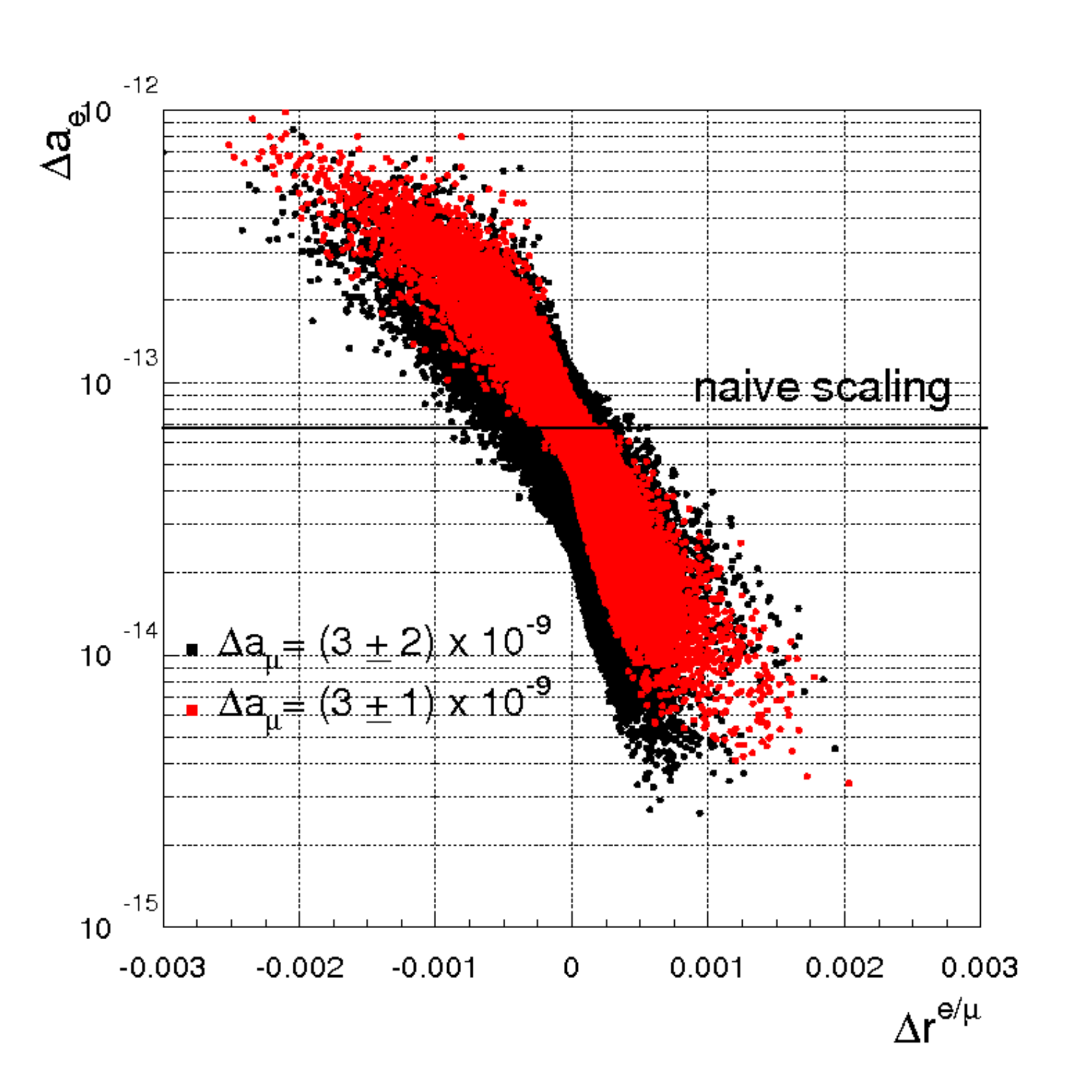}
\includegraphics[width=0.49\textwidth]{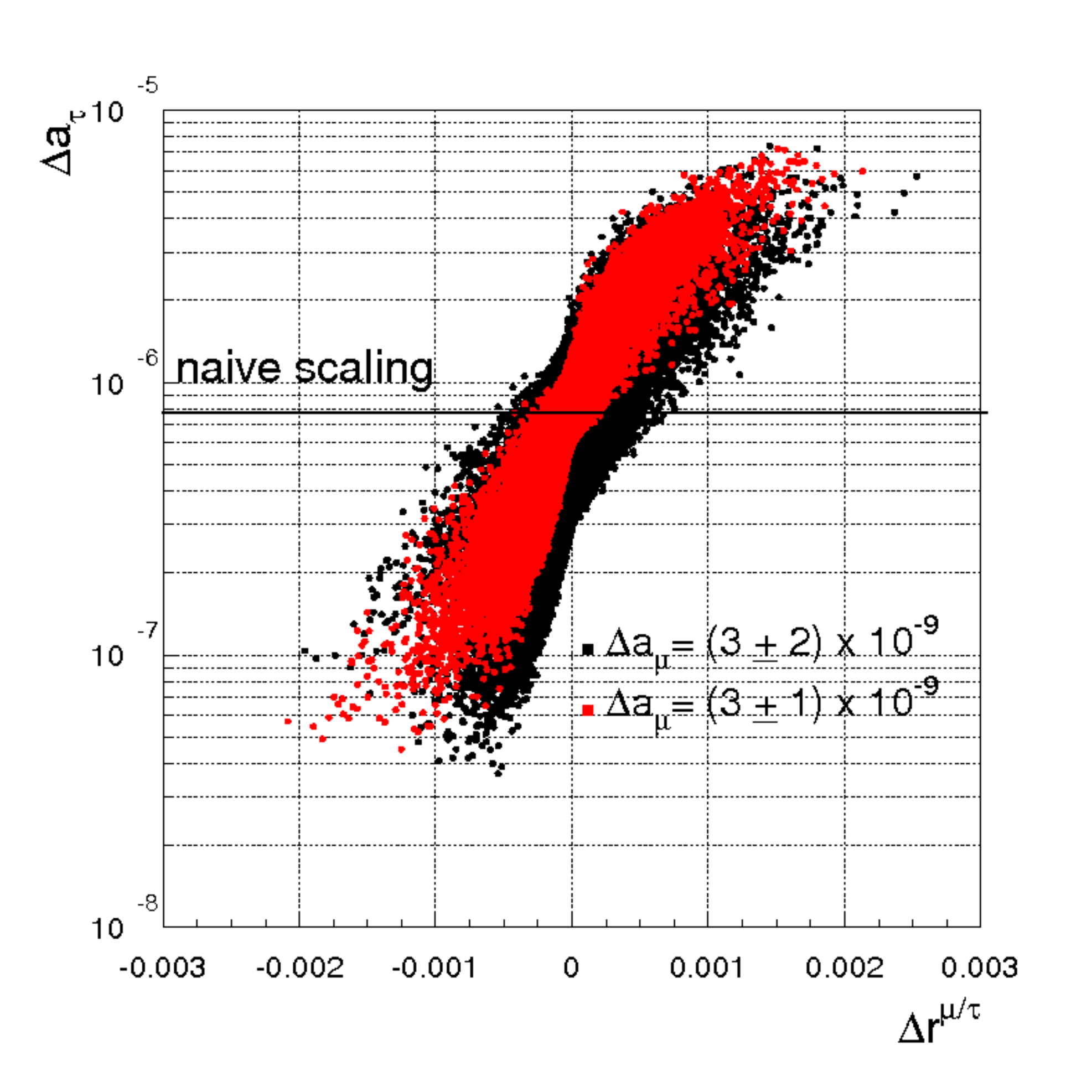}
\caption{Left: $\Delta r_{\rm P}^{e/\mu}$ as a function of $\Delta a_e$,
where $\Delta r_{\rm P}^{e/\mu}$ measures violations of lepton universality
in $\Gamma(P\to e\nu)/\Gamma(P\to \mu\nu)$ with $P=K,\pi$.
Right: $\Delta r_{\rm P}^{\mu/\tau}$ as a function of $\Delta a_\tau$ where
$\Delta r_{\rm P}^{\mu/\tau}$ measures violations of lepton universality in
$\Gamma(P\to \mu\nu)/\Gamma(\tau\to P\nu)$ with $P=K,\pi$.
Black points satisfy the condition $1\leq \Delta a_{\mu}\times 10^{9}\leq 5$,
while red points correspond to $2\leq \Delta a_{\mu}\times 10^{9}\leq 4$.}
\label{fig:LFC_emu_LFU}
\end{figure*}

In fig.~\ref{fig:LFC_emu_LFU}, on the left, which has been obtained through the
same scanning procedure described for fig.~\ref{fig:LFC}, we show the quantity
$\Delta r_{\rm K,\pi}^{e/\mu}$, defined in eq.~(\ref{RPemu}), as a function
of $\Delta a_e$. We learn that large {\small NS} violations typically
imply also breaking effects of lepton universality at the few per-mill level,
which are expected to be within the future experimental reach.

Similarly, in fig.~\ref{fig:LFC_emu_LFU}, on the right, we show breaking effects of lepton universality in the $\mu/\tau$ sector, accounted for by the quantity $\Delta r_{\rm K,\pi}^{\mu/\tau}$ which can be constructed for instance combining processes like $\tau\to P \nu$ and $P\to\mu\nu$ (where $P=\pi,K$). Large {\small NS} violations, bringing $\Delta a_{\tau}$ to the level of $10^{-5}$, are typically correlated with violations of lepton universality in the $\mu/\tau$ sector at the per-mill level.

These effects will be tested experimentally at future B factories, which aims at {\small LFU} tests in $\tau$ decays well below the per-mill level~\cite{Bona:2007qt}.
%

\subsection{Lepton flavor violating case}

We now consider {\small LFV} contributions to $a_\ell$. The importance of this case lies in the
fact that {\small LFV} can generate contributions to $a_e$ and $a_\mu$ that are chirally enhanced,
being proportional to $m_\tau$~\cite{Masina:2002mv,Girrbach:2009uy}. Using the mass-insertion approximation, which is valid for near degenerate sleptons with mass $m_{\tilde\ell}$, we find
\beq
\Delta a_\ell^{\mysmall \rm LFV}
\simeq
\frac{\alpha_1}{\pi}
\left(\frac{m_{\ell}m_{\tau}}{m^{2}_{\tilde\ell}}\right)\tan{\beta}~
\frac{{\rm Re}\left(\mu M_{1}\delta_{RR}^{\ell\tau}\delta_{LL}^{\tau\ell}\right)}{m^{2}_{\tilde\ell}}~
l_{n}(x_1)~\qquad \ell = e, \mu\,.
\label{eq:gm2_lfv}
\eeq
%
%
%
We defined $\delta_{AB}^{ij}=(m_{AB}^2)^{ij}/m^{2}_{\tilde\ell}$ (with
$A,B=L,R$), where $m_{LL}^2$ and $m_{RR}^2$ stand for the left- and right-handed
slepton mass square matrices. The function $l_n$ is given in the appendix. 
In the illustrative case where $m_{\tilde\ell}=M_1$, we find
\begin{equation}
\Delta a_\ell^{\mysmall \rm LFV}\approx 5 \times 10^{-13}
\left(\frac{m_\ell}{m_e}\right)
\left(\frac{\tan{\beta}}{30}\right)
\left(\frac{2\,{\rm TeV}}{m_{\tilde\ell}}\right)^{2}
\left(\frac{\mu/m_{\tilde\ell}}{2}\right)
\left(\frac{\delta_{RR}^{\ell\tau}}{0.5}\right)
\left(\frac{\delta_{LL}^{\tau\ell}}{0.5}\right)\,
\qquad \ell = e, \mu\,.
\label{eq:gm2_lfv2}
\end{equation}
As evident from eq.~(\ref{eq:gm2_lfv}), the single power of $m_\ell$ and the flavor
mixing angles of the soft sector break {\small NS} and provide a potentially
large chiral enhancement $m_\tau/m_{e,\mu}$, which is especially important for
$\Delta a_{e}$~\cite{Girrbach:2009uy}.

\subsubsection{Correlation between $a_{e}$ and $\tau \to e \gamma$ in LFV}

\begin{table}[t]
\centering
\vspace{0.2cm}
\renewcommand{\arraystretch}{1.2}
\begin{tabular}{@{\hskip .1cm}lll}
\hline\hline
 ${\rm BR}(\mu^-\to e^-\gamma) < 2.4 \times 10^{-12}$~\cite{MEG:11} \\
\hline
 ${\rm BR}(\tau^-\to \mu^-\gamma) < 4.4 \times 10^{-8}$~\cite{LFVbabar,LFVbelle} \\
\hline
 ${\rm BR}(\tau^-\to e^-\gamma) < 3.3 \times 10^{-8}$~\cite{LFVbabar,LFVbelle}
 \\ \hline\hline
\end{tabular}
\caption{90\% C.L.\ limits on radiative {\small LFV} decays.}
\label{table:LFV}
\end{table}

\begin{figure}
\centering
\includegraphics[width=0.49\textwidth]{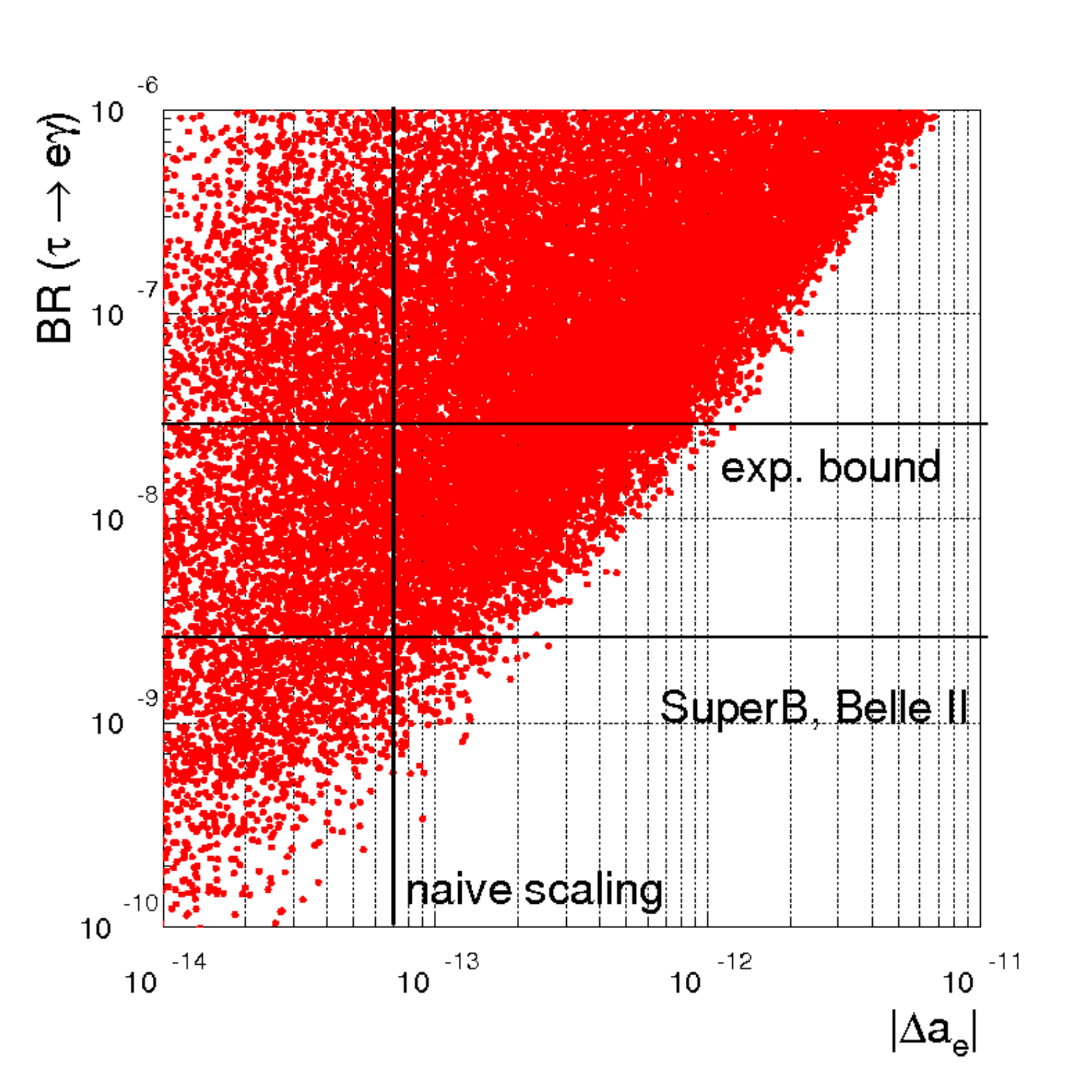}
\includegraphics[width=0.49\textwidth]{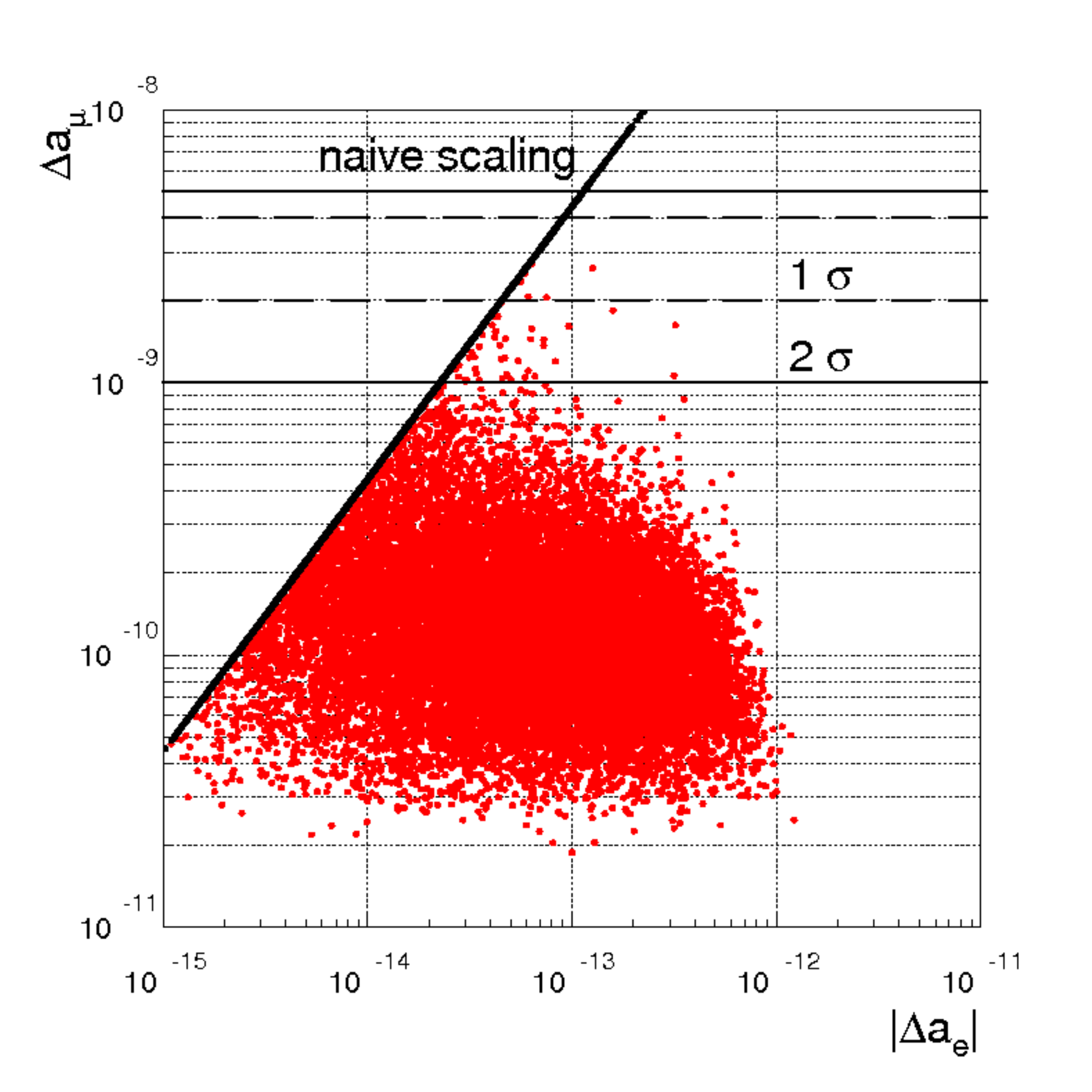}
\caption{Left: ${\rm BR}(\tau\to e\gamma)$ vs.\ $|\Delta a_{e}|$ in the {\small LFV} scenario.
The horizontal lines show the current bound and the expected future experimental
sensitivity on ${\rm BR}(\tau\to e\gamma)$. The vertical line corresponds
to the prediction for $\Delta a_{e}$ assuming a naive scaling setting
$\Delta a_{\mu}$ to its central value $\Delta a_{\mu} = 3\times 10^{-9}$.
Right: $\Delta a_{\mu}$ vs.\ $|\Delta a_{e}|$ in the {\small LFV} scenario
imposing the bound ${\rm BR}(\tau\to e\gamma) < 3.3 \times 10^{-8}$.
The black line shows the correlation between $\Delta a_{\mu}$ and
$\Delta a_{e}$ in the case of naive scaling.
The horizontal dashed (solid) lines show the 1$\sigma$ (2$\sigma$)
$\Delta a_\mu$ anomaly $\Delta a_\mu \approx (3 \pm 1) \times 10^{-9}$,
see eq.~(\ref{eq:gmu}).
}
\label{fig:teg_gm2e}
\end{figure}

The case of {\small LFV} has to be confronted with the strong constraints from processes
like $\ell_i\to\ell_j\gamma$, see table~\ref{table:LFV}. The decay rate of
$\ell_i\to\ell_j\gamma$ is given by
\begin{eqnarray}
\frac{{\rm BR}(\ell_i\to\ell_j\gamma)}{{\rm BR}(\ell_i\to\ell_j\nu_i\bar{\nu}_j)}= \frac{48\pi^3\alpha}{G^2_F}\left(|A^{ij}_L|^2+|A^{ij}_R|^2\right)\,.
\end{eqnarray}
The supersymmetric contributions to $\mu\to e\gamma$ and $\tau\to\ell\gamma$ ($\ell=e,\mu$)
amplitudes arise at one loop level through the exchange of charginos/sneutrinos and
neutralinos/sleptons. Although our numerical results are based on the exact expressions
for $A^{ij}_{L,R}$ in the mass eigenstate basis~\cite{Hisano:1995cp}, in the following
we provide their expressions in the mass-insertion approximation~\cite{Paradisi:2005fk}
\begin{eqnarray}
\label{MIamplL}
A^{ij}_L &\simeq& \frac{\alpha}{4\pi\sin^2\theta_{W}}
\frac{\delta_{LL}^{ij}}{m_{\tilde \ell}^{2}}
\frac{\mu M_{2}\tan{\beta}}{(M_{2}^2-\mu^2)}
~\bigg[g_{n}(x_{2},x_\mu) + g_{c}(x_{2},x_\mu)\bigg]
\nonumber\\
&+&
\frac{\alpha}{4\pi\cos^2\theta_{W}}
\frac{\delta_{LL}^{ij}}{m_{\tilde \ell}^{2}}
\mu M_{1}\tan{\beta}~
\bigg[
\frac{h_{n}(x_1)}{m_{\tilde \ell}^{2}} - \frac{g_{n}(x_1,x_\mu)}{(M_{1}^2-\mu^2)}
\bigg]
\nonumber\\
&+&
\frac{\alpha}{4\pi\cos^2\theta_{W}}~
\frac{\delta_{RL}^{ij}}{m_{\tilde\ell}^2}~
\left(\frac{M_1}{m_{\ell_i}}\right)~2f_{n}(x_1)~,
\end{eqnarray}
\begin{eqnarray}
\label{MIamplR}
A^{ij}_R
\simeq
\frac{\alpha}{4\pi\cos^2\theta_{W}}
\left[
\frac{\delta_{RR}^{ij}}{m_{\tilde \ell}^{2}}\mu M_{1}\tan{\beta}
\left(\frac{h_{n}(x_1)}{m_{\tilde \ell}^{2}} +
\frac{2g_{n}(x_1,x_{\mu})}{(M_{1}^2-\mu^2)}\right)
+\frac{\delta_{LR}^{ij}}{m_{\tilde\ell}^{2}}~
\left(\frac{M_1}{m_{\ell_i}}\right)~2f_{n}(x_1)
\right]~.
\end{eqnarray}
In the region of parameter space that maximizes $\Delta a_{e}$, corresponding to
moderate/large $\tan\beta$ values and a large $\mu$ term, ${\rm BR}(\tau\to\ell\gamma)$
can be approximated as
\be
{\rm BR}(\tau \to \ell\gamma) \approx
3\times 10^{-8}
\left(\frac{\tan{\beta}}{30}\right)^{2}
\left(\frac{2\,{\rm TeV}}{m_{\tilde{\ell}}}\right)^{4}
\left(\frac{\mu/m_{\tilde{\ell}}}{2}\right)^2
\left(\left|\frac{\delta_{LL}^{\tau\ell}}{0.5}\,\right|^2 +
\left|\frac{\delta_{RR}^{\tau\ell}}{0.5}\right|^2\right)\, .
\label{brtau}
\ee
Combining \eq{brtau} with eq.~(\ref{eq:gm2_lfv2}), one can set an upper bound on
$\Delta a_\ell$ as a function of ${\rm BR}(\tau\to\ell\gamma)$. We have studied
this correlation numerically and we show in fig.~\ref{fig:teg_gm2e} our results
for $\Delta a_{e}$ vs.\  ${\rm BR}(\tau\to e\gamma)$.
\begin{figure}
\centering
\includegraphics[width=0.49\textwidth]{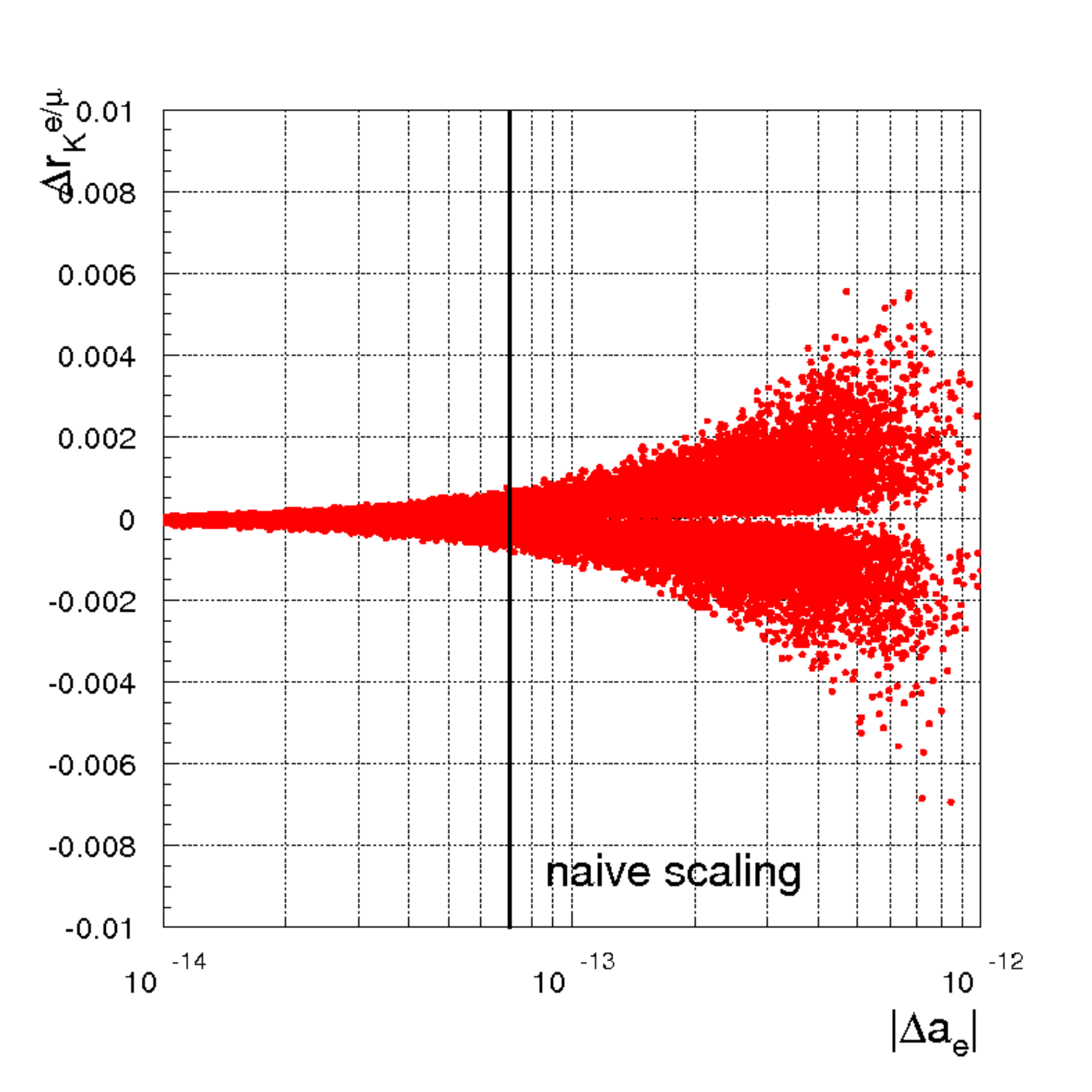}
\caption{
$\Delta r^{e/\mu}_{K}$ vs.\ $|\Delta a_{e}|$ in the {\small LFV} scenario.
The vertical line corresponds to the prediction for $\Delta a_{e}$
assuming naive scaling, setting $\Delta a_{\mu}$ equal to its central
value $\Delta a_{\mu} = 3\times 10^{-9}$.}
\label{fig:rk_gm2e}
\end{figure}
The plot has been obtained by means of a scan with $M_1 \leq 2$~TeV, $M_2 = 2 M_1$,
$\mu \leq 5$~TeV, $3 \leq\tan\beta\leq 30$, $m_{{\tilde \ell}}\leq 2$~TeV,
$|\delta_{RR}^{\tau e}|<1$, $|\delta_{LL}^{\tau e}|<1$. As illustrated in fig.~\ref{fig:teg_gm2e}, the experimental limit ${\rm BR}(\tau\to e\gamma)<3.3 \times 10^{-8}$ (90\% CL) curiously corresponds to a limit on $\Delta a_{e}$ close
to $10^{-12}$, which is just at the edge of present experimental sensitivity.
Therefore, in the case of {\small LFV}, any future positive indication for $\Delta a_{e}$
will necessarily imply that $\tau\to e\gamma$ is just beyond the present bound.
Moreover, even if $\tau\to e\gamma$ will not be detected at the upcoming SuperB
and Belle II facilities, which are expected to reach the sensitivity of
${\rm BR}(\tau\to e\gamma) \lesssim 3 \times 10^{-9}$, we can still expect values
for $\Delta a_{e}$ up to the level of $\Delta a_{e}\lesssim 2\times 10^{-13}$, which
is roughly a factor of three above the expectations of the {\small NS} scenario.

However, in contrast to the {\small LFC} scenario, {\small LFV} is not able to account for the
$\Delta a_{\mu}$ anomaly, with departure from {\small NS}. This is because by
combining \eq{brtau} with eq.~(\ref{eq:gm2_lfv2}) we find that the experimental limit
${\rm BR}(\tau\to \mu\gamma)<4.4 \times 10^{-8}$ (90\% CL) sets a very strong upper bound
on $\Delta a_\mu^{\mysmall \rm LFV}$. Moreover, large effects in $\Delta a_{e}^{\mysmall \rm LFV}$
require slepton masses roughly at the TeV, in order to keep ${\rm BR}(\tau\to e\gamma)$
under control. In such a regime, we cannot employ the contribution to $(g-2)_\mu$ from
{\small LFC} because $\Delta a_\mu^{\mysmall \rm LFC}$ is not large enough in order to account for the
anomaly, see \eq{eq:gm2_susy_1}.
This can be clearly seen in fig.~\ref{fig:teg_gm2e} on the right where we show 
$\Delta a_{\mu}$ vs.\ $|\Delta a_{e}|$ in the {\small LFV} scenario, imposing the bound
${\rm BR}(\tau\to e\gamma) < 3.3 \times 10^{-8}$. The black line shows the
correlation between $\Delta a_{\mu}$ and $\Delta a_{e}$ in the case of {\small NS} where {\small LFV} contributions are absent. In the interesting region where
$10^{-13} \lesssim \Delta a_e \lesssim 10^{-12}$, it turns out that $\Delta a_\mu$
is below the level of $10^{-9}$.

Finally, there is a formidable constraint from $\mu\to e\gamma$, a process that
receives contributions from the combination of flavor mixing angles $\delta_{\mu e}\approx\delta^{\mu \tau}\delta^{\tau e}$.
Therefore, in order to fulfill the ${\rm BR}(\mu\to e\gamma)$ bound while generating
sizable effects for $\Delta a_{e}\propto \delta^{e \tau} \delta^{\tau e}$, we need
a strong hierarchy such that $\delta_{AB}^{\mu\tau} \ll \delta_{AB}^{e\tau}$.
In conclusion, a large contribution to $\Delta a_e$ in the {\small LFV} case is possible
if the only sizable mixing is between the first and third generation of sleptons.

Also in the {\small LFV} case a potential constraint can arise from the leptonic {\small EDM}s.
In particular, in order to obtain $\Delta a_e \approx 10^{-13(12)}$ we need a suppression
for the relevant CPV phase ${\rm arg}(\mu M_{1}\delta_{RR}^{\ell\tau}\delta_{LL}^{\tau\ell})
\lesssim 10^{-3(4)}$. However, in contrast with the {\small LFC} case, we have now {\it flavor-dependent}
CPV phases and therefore the electron {\small EDM} does not constrain directly $d_{\mu,\tau}$.
Still, whenever $d_{\mu,\tau}$ are induced by {\it flavor-dependent} phases (coming
from {\small LFV} sources), powerful bounds are obtained by the {\small LFV} processes $\ell_i\to\ell_j \gamma$.
In particular, we find that the current bound ${\rm BR}(\tau\to\mu\gamma)<4.4\times 10^{-8}$
sets the upper bounds $d_{\mu}\lesssim 3\times 10^{-23}e~$cm and
$d_{\tau}\lesssim 1.5\times 10^{-24}e~$cm.

\subsubsection{Correlation between $a_{e}$ and violation of lepton universality in LFV}

Violations of lepton universality can arise also in the {\small LFV} scenario.
The quantity that is determined experimentally and accounts for deviations
from $\mu$--$e$ universality is
\beq
\left(R_P^{e/\mu}\right)_{\mysmall \rm EXP} =
\frac{\sum_i\Gamma(P\rightarrow e\nu_i)}
{\sum_i\Gamma(P\rightarrow \mu\nu_{i})}
\,\,\,\,\,\,\,\,\,\,\,\,\,\,\,\,\,\,\,\,\,\,\,\,\,\,\,\,i= e,\mu,\tau.
\eeq
The sums extend over all neutrino (or antineutrino) flavors since they cannot be
distinguished experimentally. This is important for our purposes because in the {\small LFV}
case one expects new contributions to $P \to e \nu_\tau$, a process that is detected
as breaking of lepton universality, rather than violation of lepton flavor.

One would naively expect that the {\small LFV} channels $P\to\ell_i\nu_k$ ($i\ne k$) are suppressed
compared with the {\small LFC} ones ($i= k$). An interesting exception is provided by charged Higgs mediated {\small LFV} contributions, which can be sizable for large $\tan\beta$ and which are chirally-enhanced by the factor $m_{\tau}/m_{\ell}$~\cite{Masiero:2005wr}. Indeed, the
dominant contribution to $\Delta r^{e/ \mu}_{P}$ in the case of mixing between
the first and third generation sleptons is
\begin{eqnarray}
\label{LFUlfv}
1+\Delta r^{e/\mu}_{P}\simeq
\left|1-\frac{m^{2}_{P}}{M^{2}_{H^\pm}}
\frac{m_{\tau}}{m_{e}}\Delta^{11}_{RL}\tan^3{\beta}\right|^{2}+
\bigg(\frac{m^{4}_{P}}{M^{4}_{H^\pm}}\bigg)
\bigg(\frac{m^{2}_{\tau}}{m^{2}_{e}}\bigg)
|\Delta^{31}_{RR}|^2\tan^6{\beta}.
\end{eqnarray}
The coefficients $\Delta^{11}_{RL}$ and $\Delta^{31}_{RR}$ measure the effective couplings 
$H^+{\bar \nu}_{eL} e_R$ and $H^+{\bar \nu}_{\tau L} e_R$ respectively, which are induced
at one-loop level by exchange of Bino or Bino-Higgsino and sleptons. Since these effective Yukawa interactions are of dimension four, the quantities $\Delta^{11}_{RL}$ and $\Delta^{31}_{RR}$ 
are not sensitive to the overall soft scale, hence avoiding supersymmetric decoupling.
In the region of parameter space where $\Delta a_{e}$ receives large effects,
i.e. for $\mu \gg m_{\tilde{\ell}}\sim M_1$, it turns out that\footnote{In particular,
starting from the exact expressions for $\Delta^{31}_{RR}$ and $\Delta^{11}_{RL}$ in
the mass eigenstate basis~\cite{Hisano:2008hn}, which we use in our numerical analysis,
one can find that
$\Delta^{31}_{RR}\simeq\frac{\alpha_{1}}{16\pi}\frac{\mu}{m_{\tilde{\ell}}}\delta_{RR}^{\tau e}$
and $\Delta^{11}_{RL}\simeq-\frac{\alpha_{1}}{32\pi}\frac{\mu}{m_{\tilde{\ell}}}
\delta_{LL}^{e\tau}\delta_{RR}^{\tau e}$ for $\mu\gg m_{\tilde{\ell}} \sim M_1$. Notice
that Im($\delta_{LL}^{e\tau}\delta_{RR}^{\tau e}$) is strongly constrained by the electron EDM. However, sizable contributions to $\Delta r^{e/\mu}_{P}$ can still be induced by Re($\delta_{LL}^{e\tau}\delta_{RR}^{\tau e}$).}
\begin{eqnarray}
\label{eq:LFUlfv_K}
\Delta r^{e/\mu}_{P} \approx
8 \times 10^{-3} \left(\frac{0.5~\rm TeV}{M_{H^\pm}}\right)^{2}
\left(\frac{\mu/m_{\tilde{\ell}}}{2}\right)
\left(\frac{\delta_{LL}^{e\tau}}{0.5}\right)
\left(\frac{\delta_{RR}^{\tau e}}{0.5}\right)
\left(\frac{\tan\beta}{30}\right)^{3}\,.
\end{eqnarray}
An inspection of eqs.~(\ref{eq:gm2_lfv},\ref{eq:LFUlfv_K}) reveals that {\small LFV} effects
contributing to $\Delta a_\ell^{\mysmall \rm LFV}$ and $\Delta r^{e/\mu}_{K}$ are correlated,
as shown in fig.~\ref{fig:rk_gm2e}. The plot has been obtained by means of a scan
with $10 \leq\tan\beta\leq 50$, $M_{H^+}\leq 1$~TeV, $M_1 \leq 2$~TeV, $M_2 = 2 M_1$,
$\mu \leq 5$~TeV, $m_{{\tilde \ell}}\leq 2$~TeV, and $|\delta_{AB}^{\tau e}|<1$.
It is interesting that $\Delta a_\ell^{\mysmall \rm LFV}$ and $\Delta r^{e/\mu}_{K}$ can
simultaneously reach experimentally testable values even for supersymmetric masses
beyond the {\small LHC} reach.

Turning to pion physics, one finds that $\Delta r^{e/\mu}_{\pi}\sim
(m^{2}_{\pi}/m^{2}_{k})\times \Delta r^{e/\mu}_{\!K}$. Therefore
$\Delta r^{e/\mu}_{\pi\,susy}$ is negligible after the constraints
from $\Delta r^{e/\mu}_{\!K\,susy}$ are imposed.

\subsection{Disoriented A-terms}

So far, we have investigated various supersymmetric scenarios and their capability
to account for the muon \gmt anomaly. In particular, we have analyzed the possibility
of breaking the {\small NS} among different leptonic $g$$-$$2$. As discussed in the
previous sections, both model-independently as well as in supersymmetric frameworks,
the major challenge we have to deal with, when generating large effects for the
leptonic $g$$-$$2$, is to keep under control other dipole transitions like the electron
{\small EDM} and $\mu\to e\gamma$. Yet, the correlation among $\Delta a_{\ell}$, $d_{\ell}$
and ${\rm BR}(\ell\to \ell^{\prime}\gamma)$ depends on the unknown flavor and CP
structure of new physics. Therefore, from a phenomenological perspective, $d_{\ell}$
and ${\rm BR}(\ell\to \ell^{\prime}\gamma)$ do not provide a direct bound on new
contributions to $\Delta a_{\ell}$.

However, it would be desirable to have a concrete scenario where the above conditions
are naturally fulfilled. In the following, we point out that this happens in the case
of the so-called ``disoriented A-terms'', invoked in ref.~\cite{Giudice:2012qq} to
account for the recently observed direct CP violation in charm decays $D\to KK,\pi\pi$.

The assumption of disoriented A-terms is that flavor violation is restricted to the
trilinear terms
\be
( \delta_{LR}^{ij})_{f} \sim
\frac{ A_f \theta^f_{ij} m_{f_j} }{m_{\tilde f}}~~~~f=u,d,\ell~,
\label{eq:LRgen}
\ee
where $\theta^f_{ij}$ are generic mixing angles. This pattern can be obtained when
the trilinear terms have the same hierarchical pattern as the corresponding Yukawa 
matrices but they do not respect exact proportionality.
A natural realization of this ansatz arises in scenarios with partial compositeness,
as recently pointed out in ref.~\cite{KerenZur:2012fr}.
Interestingly, the structure of eq.~(\ref{eq:LRgen}) allows us to naturally satisfy
the very stringent flavor bounds of the down-sector thanks to the smallness of down-type quark masses. Similarly, also the bounds from the lepton sector can be satisfied under
the (natural) assumption that the unknown leptonic flavor mixing angles are of the
form $\theta^{\ell}_{ij}\sim\sqrt{m_i/m_j}$~\cite{KerenZur:2012fr}.

In the disoriented A-term scenario, the dominant amplitude for $\mu\to e\gamma$ is
\be
A^{\mu e}_{L} =
\frac{\alpha~M_1~\delta_{LR}^{\mu e}}{2\pi\cos^2\theta_{W}~m_{\tilde\ell}^2~m_\mu}
~f_{n}(x_1)\,,
\label{eq:amplitude_lfv}
\ee
%
where $x_{1} = M_{1}^2/m_{\tilde\ell}^2$.
Assuming that the only possible sources of CP violation arise from A terms,
the electron {\small EDM} $d_e$ is generated by the one-loop exchange of Bino and
charged sleptons. One can find the following approximate expression
\be
\frac{d_{e}}{e} =
\frac{\alpha ~{\rm Im}\left(M_{1}\delta^{ee}_{LR}\right)}{2\pi \cos^{2}\theta_{W}
~{\tilde m}^2}~
f_{n}(x_1)\,.
\label{eq:de_susy}
\ee
On the other hand, the \gmt does not require any source of CP violation
and therefore it always receives  effects also from SU(2) interactions. In particular, the leading effects is
\be
\Delta a_\ell \simeq \frac{\alpha ~m^{2}_{\ell}~\tan\beta}{\pi \sin^{2}\theta_{W}~{\tilde m}^2}~
f^{\prime}(x_2)\,,
\label{eq:gm2_susy}
\ee
where we have considered the illustrative case where $M_{2}=\mu$, so that
$x_2 = M^{2}_{2}/{\tilde m}^2=\mu^{2}/{\tilde m}^2$. The loop function
is such that $f^{\prime}(1)= 1/8$ and $f^{\prime}(0)= 1/2$.
%
%

Therefore, the leading {\small SUSY} contribution for \gmt is parametrically enhanced
relatively to the amplitude generating $d_e$ and $\mu\to e\gamma$ by a factor of
$\sim \tan\beta/\tan^{2}\theta_{W}\approx 100 \times(\tan\beta/30)$. This naturally
raises the question of whether it is possible to account for the $(g-2)_\mu$ anomaly,
while satisfying in a natural way the constraints from $d_e$ and $\mu\to e\gamma$.
Indeed, the latter constraints generally require slepton masses around the TeV
scale, if we assume $A \sim {\tilde m}$ with $\mathcal{O}(1)$ phases and
$\theta^{\ell}_{ij} \sim \sqrt{m_i/m_j}$.
Even with such large slepton masses, it is still possible to account for the
$(g-2)_\mu$ anomaly if gauginos are significantly lighter than sleptons,
which is also welcome from naturalness arguments. This can be seen observing
that if $x_2\ll 1$, that is if $M_{2},\mu\ll {\tilde m}$, the relevant loop
function for $\Delta a_\mu$ is enhanced by a factor of four compared to the 
case where $M_{2}=\mu={\tilde m}$.
At the same time, both ${\rm BR}(\mu\to e\gamma)$ and $d_{e}$ tend to decrease
when $M_{1} \ll {\tilde m}$, see eqs.~(\ref{eq:amplitude_lfv},\ref{eq:de_susy}).
Clearly, one could equivalently consider a lighter supersymmetric spectrum to explain
the muon \gmt and $A_\ell$ terms somewhat smaller than ${\tilde m}$.

\begin{figure*}
\centering
\includegraphics[width=0.5\textwidth]{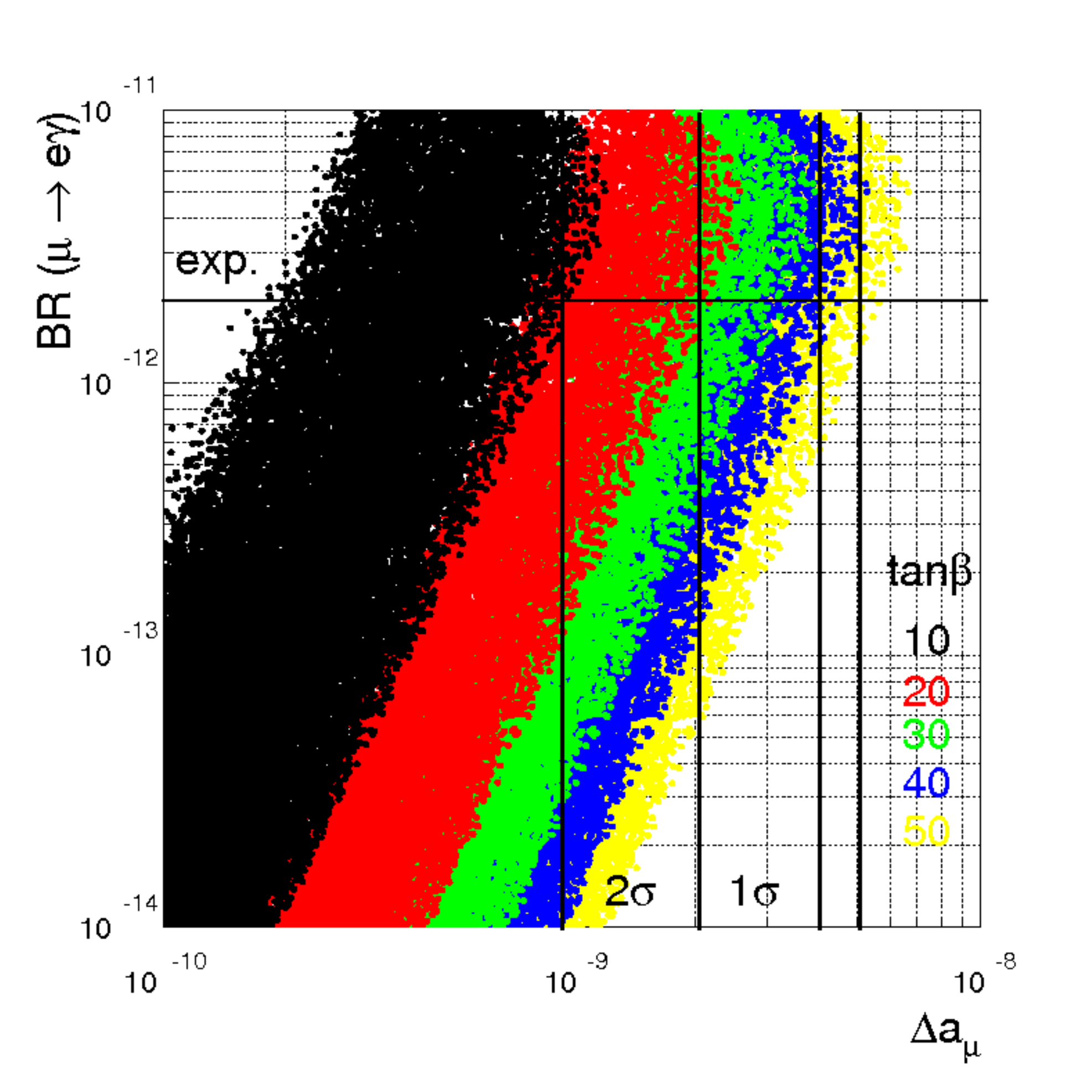}~~~
\includegraphics[width=0.5\textwidth]{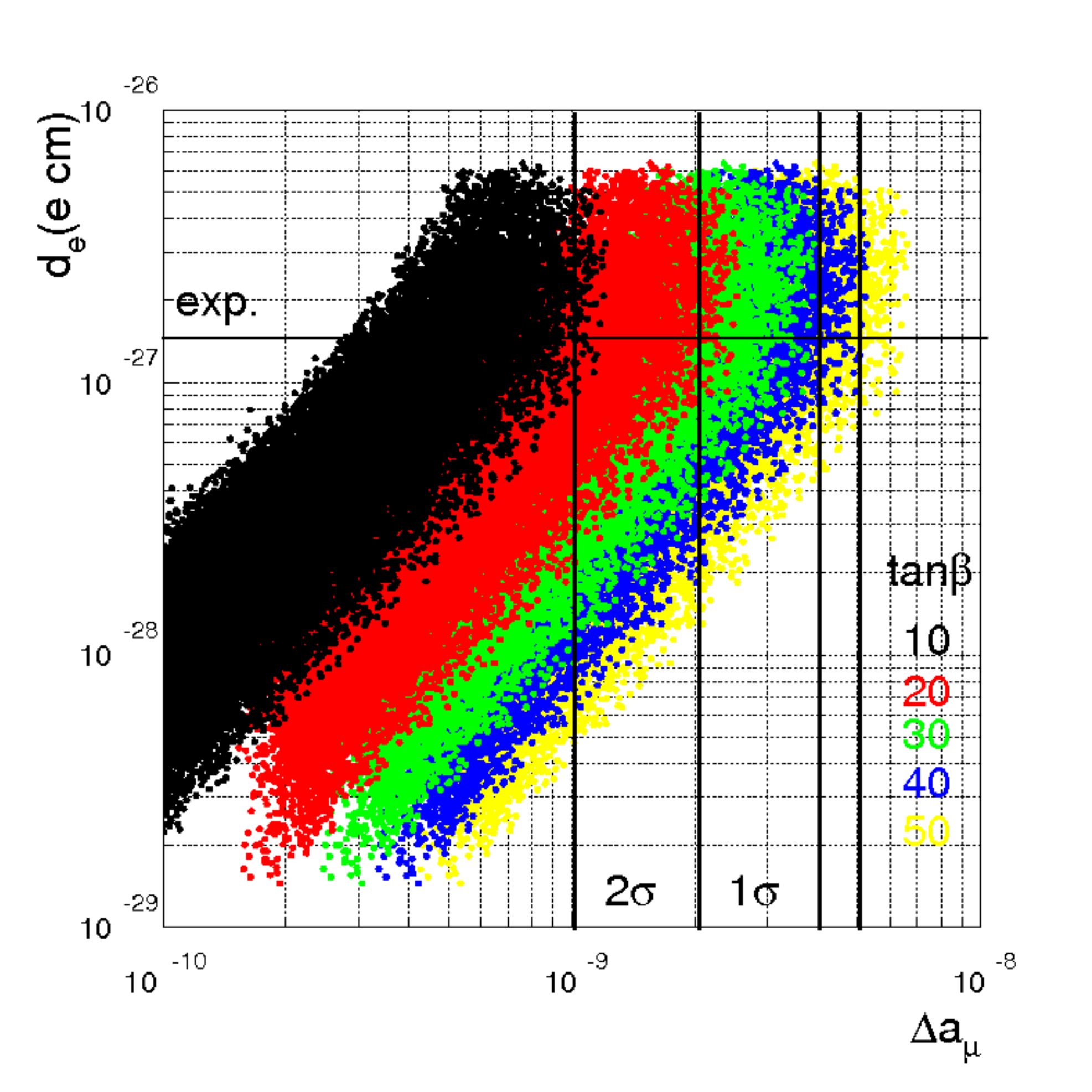}
\caption{Predictions for $\mu\to e\gamma$, $\Delta a_\mu$ and $d_e$ in the disoriented
A-term scenario~\cite{Giudice:2012qq} assuming $\theta^{\ell}_{ij}=\sqrt{m_i/m_j}$.
Left: $\mu\to e\gamma$ vs.\ $\Delta a_\mu$. Right: $d_e$ vs.\  $\Delta a_\mu$.}
\label{fig:disoriented}
\end{figure*}

In particular, setting ${\tilde m} = |A_e| =1~{\rm TeV}$, $\sin\phi_{A_e}$=1,
$M_{2} = \mu = 2M_{1} = 0.2~{\rm TeV}$, and $\tan\beta=30$, we find that
\bea
{\rm BR}(\mu\to e\gamma)
&\approx&
6\times 10^{-13}
\left| \frac{A_\ell}{\rm TeV} \frac{\theta^{\ell}_{12}}{\sqrt{m_e/m_\mu}} \right|^2
\left( \frac{\rm TeV}{m_{\tilde \ell}} \right)^4\,,\\
\nonumber
d_{e}
&\approx&
4\times 10^{-28}~
{\rm Im}\left(\frac{A_\ell~\theta^{\ell}_{11}}{\rm TeV} \right)
\left( \frac{\rm TeV}{m_{\tilde \ell}} \right)^2 e~{\rm cm}\,,\\
\nonumber
\Delta a_\mu
&\approx&
1\times 10^{-9}
\left( \frac{\rm TeV}{m_{\tilde \ell}} \right)^2
\left(\frac{\tan\beta}{30}\right)\,.
\label{eq:numeric}
\eea
These estimates are fully confirmed by the numerical analysis shown in
fig.~\ref{fig:disoriented} which has been obtained by means of the following scan:
$0.5 \leq |A_e|/{\tilde m}\leq 2$ with $\sin\phi_{A_e}$=1, ${\tilde m}\leq 2~{\rm TeV}$,
$(M_{2},\mu,M_{1}) \leq 1~{\rm TeV}$ and $10\leq \tan\beta \leq 50$.

It is interesting that disoriented A-terms can account for $(g$$-$$2)_\mu$, satisfy
the bounds on $\mu \to e\gamma$ and $d_e$, while giving predictions within
experimental reach. However, we expect that the electron \gmt follows {\small NS}.
A potential source of {\small NS} breaking comes from the trilinear terms
(provided $A_e/A_\mu \neq m_e/m_\mu$). In practice, as already discussed in
previous sections, their effects are very small after the vacuum stability
bound $|A_\ell|/m_\ell\lesssim 3$ are imposed and therefore the {\small NS} relations are preserved.

\section{Light (pseudo)scalars and $a_e$}
\label{sec:ps}

In this section we will investigate scenarios where the muon \gmt anomaly is
accounted for by contributions arising from light (pseudo)scalar
particles.\footnote{The relevance of light vector boson contributions to the
muon and electron \gmt has been recently discussed in~\cite{Davoudiasl:2012ig}.}
Irrespectively of the underlying theoretical motivations, such framework is
interesting because it typically predicts large and very special {\small NS} violations.

We parametrize the Yukawa interactions between the light scalar field $\phi$
and pseudoscalar $A$ with leptons $\ell$ with the following effective Lagrangian
\be
{\cal L} = \left(\frac{g m_\ell}{2 M_W}\right) C^{\ell}_{\phi}~\bar{\ell} \ell \phi +
i \left(\frac{g m_\ell}{2 M_W}\right) C^{\ell}_{A}~\bar{\ell}\gamma_5 \ell A\,,
\label{eq:eff_lagr_axion}
\ee
where $C^{\ell}_{\phi}$ and $C^{\ell}_{A}$ are arbitrary constants.
Although we will not discuss specific models, the field $A$ could arise as a
pseudo-Goldstone boson of an extended Higgs sector and the field $\phi$ could
be a light gauge singlet coupled through a dimension-five effective interaction
to the ordinary Yukawa terms.

Very light scalar or pseudoscalar particles with Yukawa-like couplings are generally
subject to stringent constraints both from low-energy data, such as meson decays, as well
as reactor and beam dump experiments (for a review see ref.~\cite{Ellwanger:2009dp}).
Most of these bounds disappear for $M_A \gsim 10~$GeV, which is the mass regime of
interest for us in order to account for the $(g$$-$$2)_\mu$ anomaly, as we will show below.
For simplicity, we can also consider the case in which $\phi$ and $A$ are coupled to
leptons, but not to quarks. Because of the smallness of the electron Yukawa coupling,
this case is even less constrained.

\subsection{One-loop effects}

The one-loop contribution due to (pseudo)scalar particles
to $(g$$-$$2)_\ell$ is~\cite{Leveille:1977rc},
\be
(\Delta a_\ell^{\phi A})_{\mysmall \rm 1 loop} =
\frac{g^2m_\ell^4}{32\pi^2M_W^2}
\left(
|C^{\ell}_{\phi}|^2 \frac{I^\ell_\phi}{M^{2}_{\phi}} -
|C^{\ell}_{A}|^2 \frac{I^\ell_A}{M^{2}_{A}}
\right)\,,
\label{eq:gm2_axion}
\ee
where the loop functions $I_{\phi,A}$ are
\be
I^\ell_\phi = \int_0^1 dz \frac{z^2(2-z)}{1 - z + z^2 r^{\ell}_{\phi}}\,,
\qquad
I^\ell_A  = \int_0^1 dz \frac{z^3}{1 - z + z^2 r^{\ell}_{A}}\,,
\ee
with $r^{\ell}_{\phi, A} = m_\ell^2/M^{2}_{\phi,A}$ and with asymptotic limits
\begin{align}
I^\ell_\phi = \bigg\{
	\begin{array}{ll}
        \frac{3}{2\,r} & ~~r \gg 1 \\
	-\ln r - \frac{7}{6} & ~~r \ll 1
	\end{array}\,,\qquad
I^\ell_A = \bigg\{
	\begin{array}{ll}
	\frac{1}{2\,r} & ~~r \gg 1 \\
	 -\ln r - \frac{11}{6} & ~~r \ll 1 \,.
	\end{array}
\label{eq:integral_axion}
\end{align}

Note that the one-loop pseudoscalar (scalar) effect is unambiguously negative (positive).
Therefore, at this level, a light pseudoscalar particle cannot explain the muon \gmt
anomaly as it contributes to $\Delta a_\mu$ with the wrong sign. Moreover, for
$m_\ell \gg M_{\phi,A}$ we find {\small NS}, $\Delta a_\ell \propto m_\ell^2$, see eqs.~(\ref{eq:gm2_axion},\ref{eq:integral_axion}). On the other hand, in the opposite
limit $m_\ell \ll M_{\phi,A}$, $\Delta a_\ell$ roughly scales with the fourth power of
lepton masses (apart from a mild logarithmic dependence) $\Delta a_\ell \propto m_\ell^4\ln (M_{\phi,A}/m_\ell )$, see eq.~(\ref{eq:integral_axion}).

\subsection{Two-loop effects}

Since in the $m_\ell \ll M_{\phi,A}$ regime the one loop contributions to $\Delta a_\ell$
are highly suppressed by the fourth power of lepton masses, two loop effects might be
relevant whenever we can avoid large powers of light lepton masses.
This is indeed the case for two-loop Barr-Zee type diagrams with an effective $A\gamma\gamma$
vertex generated by the exchange of heavy fermions. Here we consider only the effect of the
$\tau$ lepton, but top and bottom should also be included if $A$ (or $\phi$) are coupled to
quarks. Therefore, the  two-loop contribution is
\begin{equation}
(\Delta a_\ell^{\phi A})_{\mysmall \rm 2 loop} = -\frac{\alpha^2 }{8\pi^2\sin^2\theta_W}\frac{m_\ell^2}{M_W^2}
m_\tau^2
\left(
{\rm Re}\left(C^{\ell}_{\phi}C^{\tau *}_{\phi}\right) \frac{L^\tau_\phi}{M_\phi^2} -
{\rm Re}\left(C^{\ell}_{A} C^{\tau *}_{A}\right) \frac{L^\tau_A}{M_A^2}
\right)\ ,
\label{eq:barr_zee}
\end{equation}
where the loop functions are
%
%
%
%
\begin{equation}
L_\phi^\tau =
\int^1_0 dz~\frac{1-2z(1-z)}{z(1-z)-r^\tau_\phi} \ln\frac{z(1-z)}{r^\tau_\phi}\,,
\qquad
L_A^\tau=\int^1_0 dz~\frac{1}{z(1-z)-r^\tau_A}\ln\frac{z(1-z)}{r^\tau_A}\,,
\end{equation}
with asymptotic limits
\begin{align}
L^\tau_\phi = \bigg\{
\begin{array}{ll}
\frac{6\ln r + 13}{9\,r} &~~r \gg 1 \\
\ln^2 r + 2\ln r +4 + \frac{\pi^3}{3} &~~r \ll 1
\end{array}\,,\qquad
L^\tau_A = \bigg\{
	\begin{array}{ll}
	\frac{ \ln r +2}{r}  &~~r \gg 1 \\
	\ln^2 r + \frac{\pi^3}{3} &~~r \ll 1 \,.
	\end{array}
\label{eq:2loop_function}
\end{align}
As shown by eq.~(\ref{eq:barr_zee}), two loop effects for $\Delta a_\ell$
exhibit {\small NS} and can be positive or negative depending on the sign
of ${\rm Re}(C^{\ell}C^{\tau *})$.
Moreover, the enhancement factor $m^2_\tau/m^{2}_{e,\mu}$ of two loop effects
relative to one-loop effects, which is particularly important for the case
of the electron $g$$-$$2$, can easily compensate the additional loop suppression,
as we will see later. On the other hand, in the case of the $\tau$, one-loop
effects are typically dominant compared to two-loop effects.

Hereafter, we will focus on the case of pseudoscalar particles, which is especially
interesting. Specializing to the mass regime $m_{\ell}\ll M_A$ where the $\Delta a_\mu$
anomaly can find an explanation, we have the following situation:
1) $\Delta a_e$ is always dominated by two-loop effects,
2) $\Delta a_\mu$ receives comparable one- and two-loop contributions,
and 3) $\Delta a_\tau$ is always dominated by one-loop effects.
As a result, we expect significant {\small NS} violations
that we are going now to study numerically.

\begin{figure*}
\centering
\includegraphics[width=0.49\textwidth]{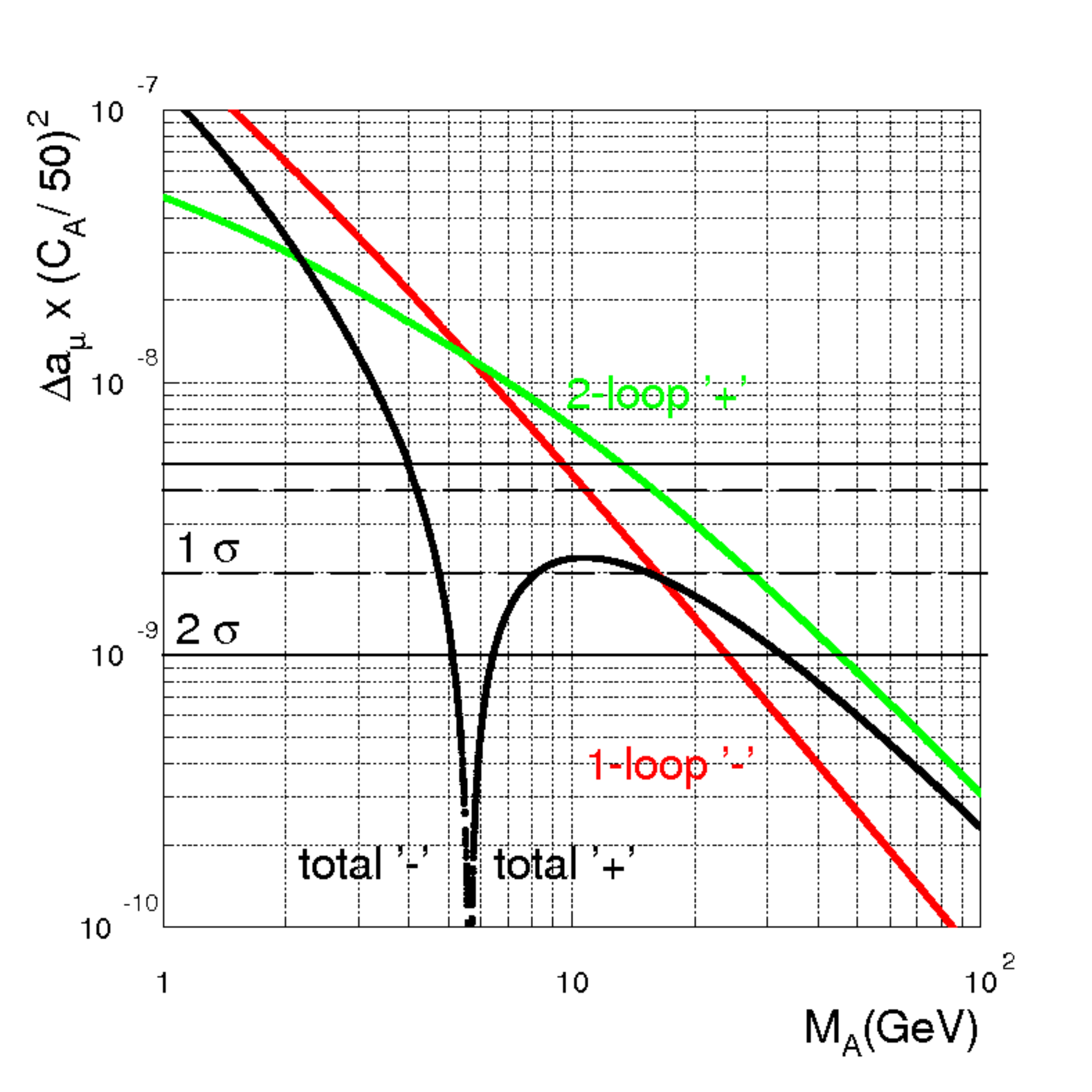}
\caption{The different contributions to $\Delta a_\mu$ induced by
a pseudoscalar particle (with mass $M_A$) as a function of $M_A$.
The red line corresponds to the negative (``$-$'') one-loop contribution,
the green line to the positive (``$+$'') two-loop contribution and
the black line to the total effect.}
\label{fig:axion}
\end{figure*}

In the left plot of fig.~\ref{fig:axion}, we show the anatomy of the
contributions to $\Delta a_\mu$ setting $C_A = 50$: the red line corresponds
to the magnitude of the negative (``$-$'') one-loop effects, the green line refers to the
positive (``$+$'') two-loop effects and the black line stands for the magnitude of
the total contribution.
The most prominent feature emerging from this plot is the different
decoupling properties of the one- and two-loop effects as one can check
directly from eq.~(\ref{eq:2loop_function}). In particular,
two-loop effects have a much milder decoupling with $M_A$ compared to one
loop-effects. This implies, in turn, the existence of a value for $M_A$
where the two effects have comparable (and opposite) size. The exact value
of $M_A$ where this happens depends on the masses and couplings of the
particles circulating in the second loop of the Barr-Zee diagram.
In our case, an almost exact cancellation occurs for $M_A \approx 6~$GeV
and therefore the total contribution to $\Delta a_\mu$ is positive for
$M_A \gtrsim 6~$GeV and negative for $M_A \lesssim 6~$GeV.

\begin{figure*}
\centering
\includegraphics[width=0.49\textwidth]{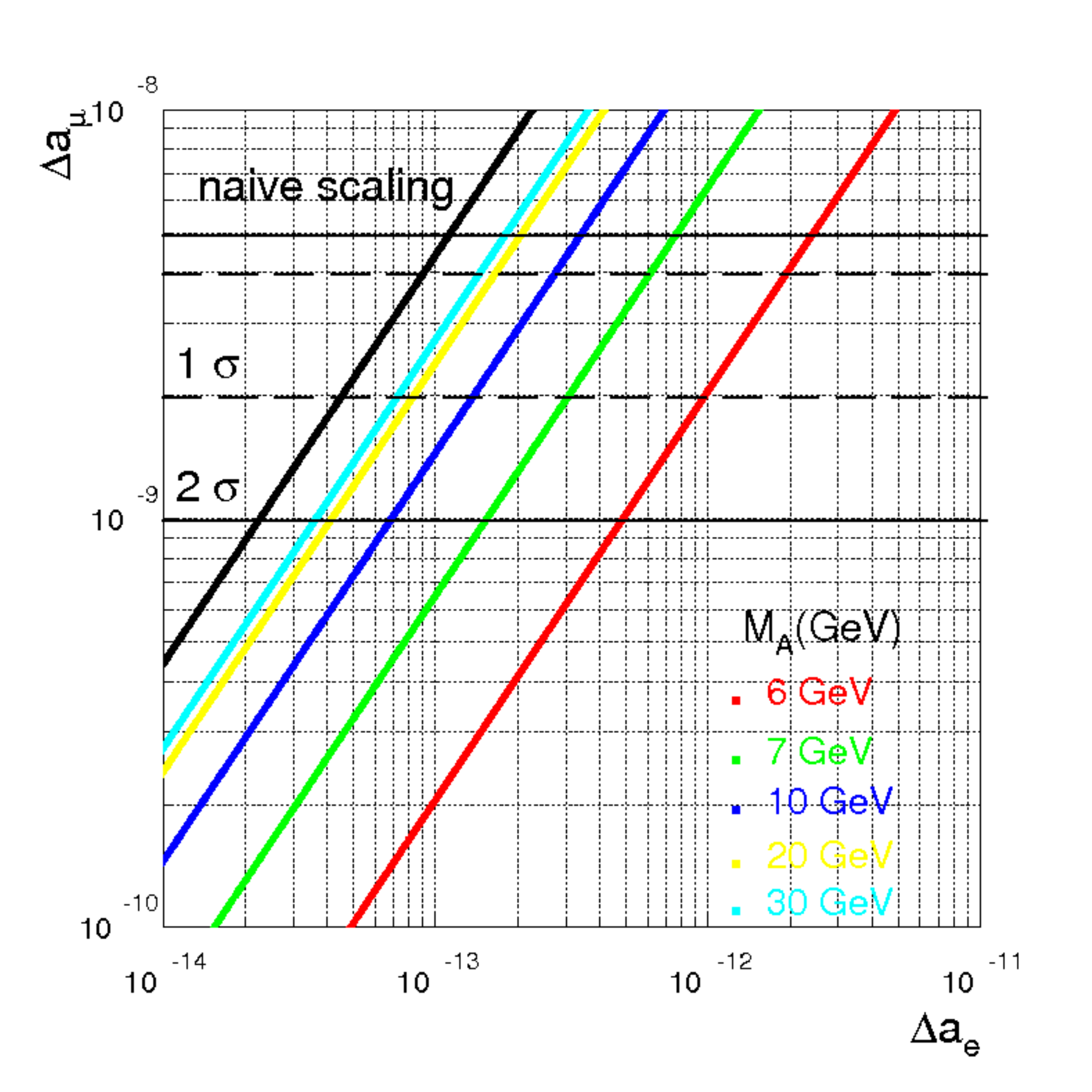}
\includegraphics[width=0.49\textwidth]{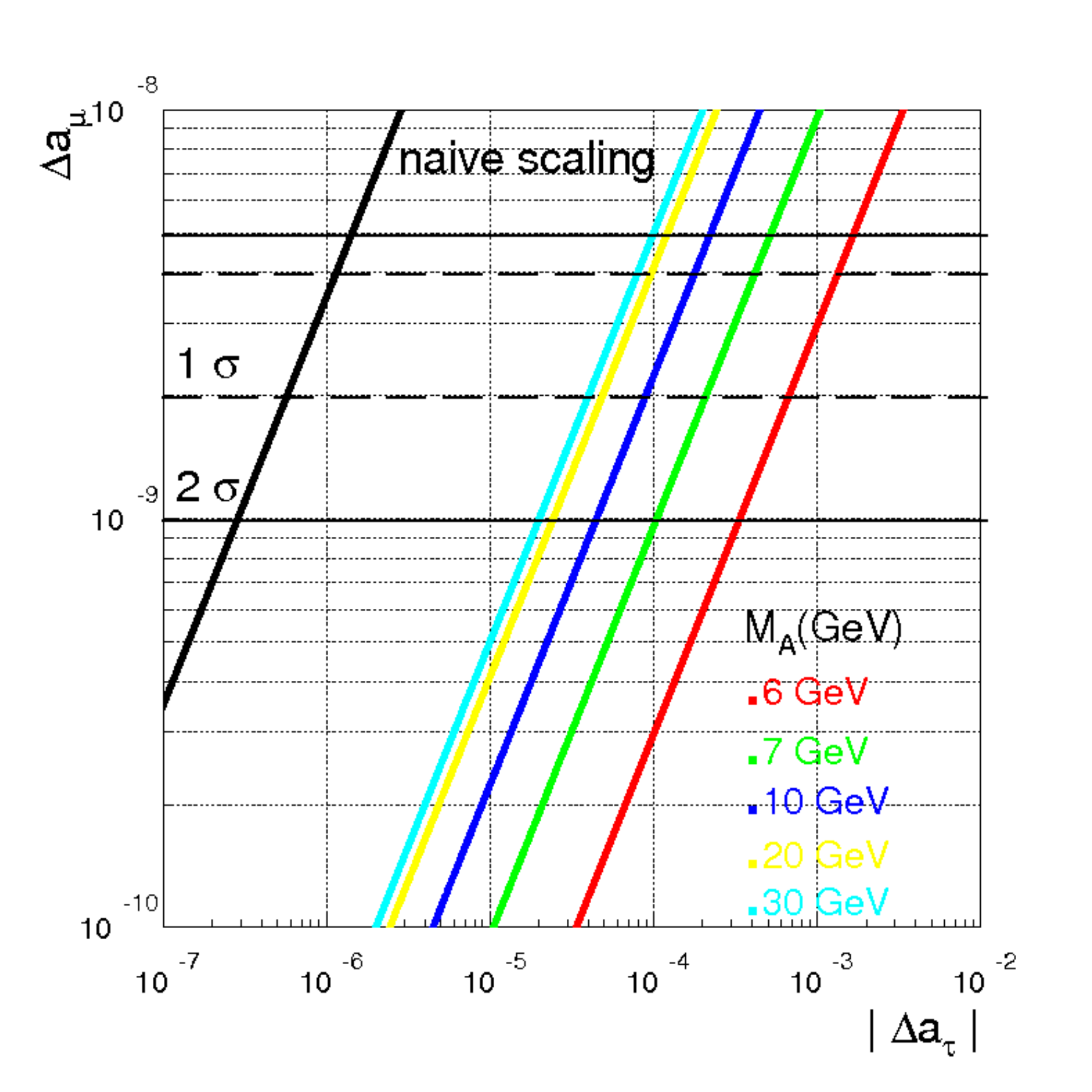}
\caption{
Left: $\Delta a_\mu$ vs.\ $\Delta a_e$ for different values of $M_A$
varying $C_A$.
Right: $\Delta a_\mu$ vs.\ $\Delta a_\tau$ for different values of $M_A$
varying $C_A$. The naive scaling predictions are given by black lines.}
\label{fig:axion_tau}
\end{figure*}

In fig.~\ref{fig:axion_tau} we show the values attained by $\Delta a_\ell$
induced by pseudoscalar effects monitoring, in particular, whether {\small NS} is at work or not.

In the left plot of fig.~\ref{fig:axion_tau}, we show $\Delta a_\mu$ vs.\
$\Delta a_e$ for different pseudoscalar masses $M_A$ and varying $C_A$.
As we can see, {\small NS} (black line) is systematically
violated by a large amount and, in particular, $\Delta a_e$ always
lies above its naive expectation.
The actual amount of {\small NS} violations depends on the degree
of cancellation between one- and two-loop effects entering $\Delta a_\mu$.
Overall, in the regions where the $\Delta a_\mu$ anomaly is accommodated,
$\Delta a_e$ typically exceeds the $10^{-13}$ level, thereby providing
a splendid opportunity to test the \gmt anomaly via the electron one.

In fig.~\ref{fig:axion_tau}, on the right, we show $\Delta a_\mu$ vs.\ $\Delta a_\tau$
for different pseudoscalar masses $M_A$ and varying $C_A$. Similarly to
the case of the electron $g$$-$$2$, {\small NS} (black line) is largely
violated and $\Delta a_\tau$ can reach values up to the level of $10^{-3}$,
while explaining the $\Delta a_\mu$ anomaly. Such values should be well
within the experimental resolutions expected at a SuperB factory.

Let us discuss now the predictions for the leptonic {\small EDM}s induced by light pseudoscalars.
At one-loop level, the effective Lagrangian of \eq{eq:eff_lagr_axion} leads to a real
dipole amplitude and therefore the {\small EDM}s are vanishing at this order.
Two loop effects are generally complex and provide very important contributions, as
already discussed for the case of the leptonic $g$$-$$2$. Since two-loop effects follow {\small NS},
it is expected that the electron {\small EDM} is the most sensitive probe of this scenario among the
leptonic {\small EDM}s. In particular, from the {\it model-independent} expectations of \eq{eq:gm2_vs_edm},
we deduce that, in order to accommodate the $(g-2)_\mu$ anomaly while satisfying the electron
{\small EDM} bounds, we need the condition ${\rm Im}(C^{\ell}C^{f*})\lesssim 10^{-3}$, where $f$ stands
here for the heavy fermion running in the second loop. This condition could be naturally
satisfied either if ${\rm Im}(C^{\ell}) = {\rm Im}(C^{f})$ or if a dynamical mechanism
suppresses both ${\rm Im}(C^{\ell})$ and ${\rm Im}(C^{f})$, for instance by a loop factor.

\section{Vector-like fermions and $a_e$}
\label{sec:vec}

In the following, we will consider the impact of heavy vector-like fermions on the
leptonic $(g-2)$ (see also~\cite{Kannike:2011ng}). The introduction of heavy vector-like fermions, mixing with the {\small SM}  fermions, can be motivated by the explanation
of flavor hierarchies in the {\small SM}, as we will discuss shortly.

Hereafter, we will focus on the leptonic sector only and therefore we introduce SU(2)
vector-like doublets ($L_L\oplus L_R$) and singlets ($E_L\oplus E_R$) governed by the
Lagrangian
\bea
\label{eq:eff_lagr1}
- {\cal L} \!\!\!&=&\!\!\! 
M_E \bar{E}_{L} E_{R} + M_L \bar{L}_{R} L_{L} + m_E \bar{E}_{L} e_{R} +
m_L \bar{L}_{R} \ell_{L} \nonumber \\
&+&
\lambda_{LE} \bar{L}_{L} E_{R} H + {\bar\lambda}_{LE} \bar{L}_{R} E_{L} H^\dagger
+ {\rm h.c.}\,.
\eea
The fields $L_L$ ($L_R$) and $E_R$ ($E_L$) have the same (opposite) quantum
numbers as the {\small SM} fields $\ell_L$ and $e_R$, respectively.
Note that the {\small SM} Yukawas are assumed to be vanishing and they will
be generated dynamically after electroweak symmetry breaking through
the mixing between light and heavy fermions, once heavy fermions are
integrated out.

The above Lagrangian is reminiscent of the fermionic sector of composite Higgs models
which are dual of warped 5d models. In such a scheme, the chiral fermions correspond
to weakly-coupled elementary fields while the vector-like fermions are composite
fields belonging to a strongly interacting sector. The Higgs boson also belongs
to the strong sector and does not couple to elementary fields.

The mass eigenstates, before electroweak symmetry breaking, are obtained diagonalizing
the mass mixing in ${\cal L}$ through the following $2 \times 2$ unitary matrices
\begin{equation}
\begin{pmatrix} \ell_L \\ L_L \end{pmatrix} \rightarrow
\begin{pmatrix} \cos\theta_{L} & \sin\theta_{L} \\
 -\sin\theta_{L} & \cos\theta_{L} \end{pmatrix}
\begin{pmatrix} \ell_L \\ L_L \end{pmatrix},~~
\begin{pmatrix} e_R \\ E_R \end{pmatrix} \rightarrow
 \begin{pmatrix} \cos\theta_{R} & \sin\theta_{R} \\
 -\sin\theta_{R} & \cos\theta_{R} \end{pmatrix}
\begin{pmatrix} e_R \\ E_R \end{pmatrix}\,,
\label{eq:mass_matrix}
\end{equation}
where
\begin{equation}
\tan\theta_{L} = \frac{m_L}{M_L}\,,\qquad
\tan\theta_{R} = \frac{m_E}{M_E}\,.
\end{equation}
Hereafter we use the notation $s_{L(R)}=\sin\theta_{L(R)}$ and
$c_{L(R)}=\cos\theta_{L(R)}$.
After performing the rotations of eq.~(\ref{eq:mass_matrix}),
the Lagrangian of eq.~(\ref{eq:eff_lagr1}) becomes
\begin{eqnarray}
-{\cal L} &=&
M^{\prime}_E \bar{E}_{L} E_{R} + M^{\prime}_L \bar{L}_{R} L_{L} +
{\bar\lambda}_{LE}\, \bar{L}_R E_L H^\dagger
\nonumber\\
&+&
\lambda_{LE}\, \left( s_L s_R~\bar{\ell}_L e_R  +
c_L s_R \bar{L}_L e_R + s_L c_R \bar{\ell}_L E_R+c_L c_R~\bar{L}_L E_R
\right) H 
+ \text{h.c.}\,,
\end{eqnarray}
where
\begin{equation}
M^{\prime}_L = \sqrt{M^2_L + m^2_L}\,,\qquad
M^{\prime}_E = \sqrt{M^2_E + m^2_E}\,,
\end{equation}
and therefore we can define the following mass mixing matrix
\beq
\label{eq:ML'E'}
M_\pm =
\bordermatrix{ & e_R & L_{R} & E_R \cr \ell_L & \lambda_{LE} s_L s_R v & 0 &
\lambda_{LE} s_L c_R v \cr L_L & \lambda_{LE} c_L s_R v &  M^{\prime}_L &
 \lambda_{LE}c_L c_Rv\cr E_L & 0 & \bar\lambda_{LE} v & M^{\prime}_E}\,.
\eeq
As a result, the fermionic spectrum consists of two heavy fermions with masses
approximatively given by $M^{\prime}_E$ and $M^{\prime}_L$ and light {\small SM}
fermions with masses given by
\begin{eqnarray}
m_{\ell_i} \simeq \lambda_{LE}\, s^i_L s^i_R \,v+ \mathcal{O}\left( \frac{v^2}{M^2_{L,E}} \right)\,,
\label{eq:lepton_masses}
\end{eqnarray}
where $i=1,2,3$ is a lepton flavor index. The leptonic $g$$-$$2$ are generated at the loop-level by the exchange of the Higgs, $W$, $Z$ bosons and heavy fermions. The dominant contribution arises from the diagram with an underlying mixing of SU(2) doublets and singlets (since it is chirally enhanced by a
factor of $v/m_\ell$).

Using the mass relation of \eq{eq:lepton_masses}, neglecting terms of order
$\mathcal{O}(v^4 / M^4_{L,E})$ and performing an explicit loop calculation
one can find that the dominant contribution is given by~\cite{Kannike:2011ng}
\begin{eqnarray}
\Delta a_\ell \simeq
\frac{c}{16\pi^2}\frac{m^2_\ell}{M_L M_E}
{\rm Re}(\lambda_{LE}{\bar\lambda_{LE}}\, c_Lc_R)
\approx c \times 10^{-9}~{\rm Re}(\lambda_{LE}{\bar\lambda_{LE}})\,
\left(\frac{300~{\rm GeV}}{\sqrt{M_L M_E}}\right)^2
\frac{m^2_\ell}{m^2_\mu}
\,.
\label{eq:gm2_vlf}
\end{eqnarray}
where $c\sim \mathcal{O}(1)$.
From eq.~(\ref{eq:gm2_vlf}) we learn that if the heavy leptons have masses around the
{\small EW} scale $M_L,M_E \sim v$ and if ${\rm Re}(\lambda_{LE}{\bar\lambda_{LE}})\sim 1$
then the muon $g$$-$$2$ anomaly can be solved. We stress that, in general, the parameters
$\lambda_{LE}$, ${\bar\lambda_{LE}}$, $M_{L,E}$, and $c_{L,R}$ are not flavor
universal and therefore naive scaling might be violated.

The Yukawa couplings $\lambda_{LE}$ and ${\bar\lambda_{LE}}$ could also violate lepton
flavor and CP. As a reference framework, we consider the so-called anarchic scenario,
where all the entries in $\lambda_{LE}$ and ${\bar\lambda_{LE}}$ are assumed to be of
order one. For the branching ratio of $\mu\to e\gamma$ we obtain
\be
{\rm BR}(\mu\to e\gamma) \approx c^2 \times 10^{-6}
|\lambda_{LE}{\bar\lambda_{LE}}|^2
\left(\frac{300\rm GeV}{\sqrt{M_LM_E}}\right)^4
\left(\frac{\left|s_{L}^{e}/s_{L}^{\mu}\right|^2 +
\left|s_{R}^{e}/s_{R}^{\mu}\right|^2}{m_e/m_\mu}
\right)\,,
\ee
and therefore, in order to satisfy the experimental constraint
${\rm BR}(\mu\to e\gamma)\lesssim 2 \times 10^{-12}$, we need
$|s^{e}_{L(R)} / s^{\mu}_{L(R)}|^2 \lesssim 2 \times 10^{-6} \frac{m_e}{m_\mu}$
for $c \sim 1$.
Such a suppression for the flavor mixing angles could be achieved
by introducing a flavor symmetry determining the structures of
$\lambda_{LE}$ and ${\bar\lambda_{LE}}$.

For the electron {\small EDM} we find
\be
\frac{d_{e}}{e} \approx 5c \times 10^{-25}~
{\rm Im}(\lambda_{LE}{\bar\lambda_{LE}})
\left(\frac{300\rm GeV}{\sqrt{M_LM_E}}\right)^2~
e~{\rm cm}\,,
\label{eq:de}
\ee
and the experimental bound is satisfied for
$|{\rm Im}(\lambda_{LE}{\bar\lambda_{LE}})|\lesssim 2 \times 10^{-3}$ for $c \sim 1$.
Thus, we conclude that both the flavor and CP structures of $\lambda_{LE}$
and ${\bar\lambda_{LE}}$ have to be highly non-generic to satisfy the
constraints from $\ell_i\to\ell_i\gamma$ and the electron {\small EDM}.

Finally, we mention that interactions with heavy leptons generate also
corrections to the fermion couplings of the $Z$ and Higgs bosons. Indeed,
after integrating out the heavy states using their equations of motion
we obtain dimension-six operators leading to corrections of the fermion
couplings of the form $v^2/M^2_{L,E}$.

\section{Conclusions}
\label{sec:conc}

The aim of this paper was to show that the anomalous magnetic moment of the electron $a_e$
can be viewed today as a new player among the low-energy processes that are able to probe 
new-physics effects. This novel status of $a_e$ stems from recent improvements on both
the experimental and theoretical fronts. One important ingredient is the measurement
of $\alpha$ from atomic-physics experiments, which are becoming competitive with $a_e$
in the determination of the fine-structure constant. The second ingredient is the ongoing
effort to measure $a_e$ with better experimental accuracy. The third element is a more
precise theoretical determination of $a_e$ in the {\small SM}. At present, the experimental measurement of $a_e$ is in good agreement with the {\small SM} and the uncertainty in
the quantity $\Delta a_e = a_e^{\rm \mysmall EXP} -a_e^{\rm \mysmall SM}$ is about
$8 \times 10^{-13}$. As discussed in this paper, future progress can reduce this error
by about one order of magnitude.

From the theoretical point of view, the great interest in testing new-physics effects in
$a_e$ comes from the well-known discrepancy between the experimental measurement and the
{\small SM} prediction of $a_\mu$. Observing or excluding an anomaly in $a_e$ could become the most convincing way to establish the origin of the $a_\mu$ discrepancy. In a large class of
models, new-physics contributions to $a_\ell$ (for $\ell =e,\mu, \tau$) are proportional
to $m_\ell^2$, a situation that we call ``naive scaling". In the case of naive scaling,
the present value of the $(g$$-$$2)_\mu$ anomaly, see \eq{eq:gmu}, corresponds
to $\Delta a_e =(0.7\pm 0.2) \times 10^{-13}$. While theoretical predictions have recently achieved an impressive precision ${\mathcal O} (10^{-13})$ via formidable calculations, confirmation of the $a_\mu$ discrepancy through $a_e$ still requires experimental improvements at the utmost level.

In many well-motivated cases, new physics effects do not respect naive scaling, as we have shown with several examples of such theories. In the context of supersymmetry, this can happen with lepton flavor conservation (for non-degenerate but aligned sleptons) or with lepton flavor violation (with relative misalignment between lepton and sleptons). In the first case, once we normalize $\Delta a_e$ in such a way to reproduce $\Delta a_\mu \approx 3\times 10^{-9}$, we obtain specific predictions on violation of lepton universality in $P \to \mu \nu /P \to e \nu$ and $P \to \mu \nu /\tau \to P \nu$, for $P=\pi , K$. The case of lepton flavor violation cannot explain the anomaly in $a_\mu$, but can lead to new effects in $a_e$ that are correlated with the prediction of $\tau \to e \gamma$ and violation of $e/\mu$ lepton universality. In a fairly model-independent way, it is also possible to correlate new effects in $\Delta a_e$ with the corresponding electric dipole moment.

We have also considered a class of theories with a light (pseudo)scalar interacting with matter proportionally to ordinary Yukawa couplings. In this case we have found an interesting pattern of violations of naive scaling coming from the interplay between one-loop contributions with $\Delta a_\ell \propto m_\ell^4$ and two-loop contributions with  $\Delta a_\ell \propto m_\ell^2$. For the parameters capable of explaining the $(g$$-$$2)_\mu$ anomaly, we found that $\Delta a_e$ is larger than its naive-scaling value and can be close to its present experimental bound. 

We believe that $a_e$ offers a special opportunity to test new-physics effects and to shed new
light on the current discrepancy in the muon anomalous magnetic moment. A robust and ambitious experimental program is necessary to exploit its full potential.

\section*{Acknowledgments}

We would like to thank L.~Mercolli for participating in the early stages of this work.
We are also grateful to G.\ Gabrielse, T.\ Kinoshita, W.\ J.\ Marciano, F.\ Nez, D.\ Straub, A.\ Strumia,
and S.\ K.\ Vempati for fruitful discussions and correspondence.
M.P.\ also thanks the Department of Physics of the University of Padova for its support.
His work was supported in part by the Italian Ministero dell'Universit\`a e della Ricerca Scientifica under the {\small COFIN} program {\small PRIN} 2008, and by the European
Programmes {\small UNILHC} (contract {\small PITN-GA}-2009-237920) and {\small INVISIBLES} (contract {\small PITN-GA}-2011-289442).

\section*{Appendix}

The loop functions entering the supersymmetric contributions
$\Delta a_\ell^{\mysmall \rm LFC}$, 
$\Delta a_\ell^{\mysmall \rm LFV}$,
and ${\rm BR}(\ell_i\to \ell_j\gamma)$ are given by
\begin{equation}
f_{n}(x) = \frac{1 - x^2 + 2x\log x}{(1-x)^3}~,
\end{equation}
\begin{equation}
f_{c}(x) = \frac{3 -4x + x^2 + 2\log x}{(x-1)^3}~,
\end{equation}
\begin{equation}
g_{n}(x) = \frac{1 + 4x -5x^2 + 2x(x+2)\log x}{4(1-x)^4}~,
\end{equation}
\begin{equation}
g_{c}(x) = \frac{5 - 4x -x^2 + 2(2x+1)\log x}{2(1-x)^4}~,
\end{equation}
\begin{equation}
h_{n}(x) = \frac{1+9x-9x^2-x^3+6x(x+1)\log x}{3(1-x)^5}~,
\end{equation}
\beq
l_{n}(x) = \frac{3 + 44x - 36x^2 - 12x^3 + x^4 + 12x(3x+2)\log x}{6(1-x)^6}~,
\eeq
We have also defined $f_{(c,n)}(x,y)=f_{(c,n)}(x)-f_{(c,n)}(y)$.



\end{document}